\documentclass[prc,preprint,showpacs,amsmath,amssymb,floatfix]{revtex4-1}
\usepackage{graphicx,color,dcolumn,booktabs,bm}

\begin{document}

\title{$\eta$ production off the proton in a Regge-plus-chiral quark approach}
\author{Jun He}
\email[]{junhe@impcas.ac.cn} 

\affiliation{Institute of Modern Physics,Chinese Academy of Sciences, Lanzhou 730000, P. R. China}

\affiliation{Research Center for Hadron and CSR Physics,Lanzhou
University $\&$ Institute of Modern Physics of CAS, Lanzhou 730000, China}

\author{B. Saghai}
\email[]{bijan.saghai@cea.fr} \affiliation{Institut de Recherche sur les lois
Fondamentales de l'Univers,
DSM/Irfu, CEA/Saclay, 91191 Gif-sur-Yvette, France}
\date{\today}
\begin{abstract}

A chiral constituent quark model approach, embodying {\it s}- and {\it u}-channel exchanges,
complemented with a Reggeized treatment for {\it t}-channel is presented.
A model is obtained allowing data for $\pi^- p \to \eta n$ and
$\gamma p \to \eta p$ to be describe satisfactorily.
For the latter reaction, recently
released data by CLAS and CBELSA/TAPS Collaborations in the system total energy range
$1.6 \lesssim W \lesssim 2.8$ GeV  are well reproduced due to the inclusion of
Reggeized trajectories instead of simple $\rho$ and $\omega$ poles.
Contribution from ``missing'' resonances, with masses below 2~GeV, is found to be negligible in the considered
processes.

\end{abstract}
\pacs{13.60.Le, 12.39.Fe,12.39.Jh,14.20.Gk}

\maketitle

\section{Introduction}\label{Sec:Intro}

The $\eta$ production offers an appropriate frame to study the properties of the nucleon
resonances ($N^*$).
The zero isospin of $\eta$ is a good filter to select the produced 1/2
isospin resonances compared with other meson production channels, such
as $\pi N$, where isospin 3/2 resonances also intervene.
Among some twenty $N^*$s, the lightest $S_{11}$ resonance, $N(1535)$, attracts much
attention, because it is related to many interesting issues on the internal structure of nucleon resonances,
such as the mixing angles
~\cite{Isgur1977,Isgur2000a,Saghai2001,Chizma2003,He2003a,He2003b,Capstick2004a,Saghai2010}
and multiquark components~\cite{An2009,An2009a}.
The substantial branching ratio of the $N(1535) \to \eta N$ decay channel still remains
to be fully understood, but it facilitates measurements of the $\eta$ production processes,
given that from threshold up to roughly 200 MeV above, the reaction mechanism is
dominated by $N(1535)$.
Moreover, those reactions offer the possibility of also investigating the properties of other
nucleon resonances and searching for still undiscovered ones.

For half a century, hadronic and electromagnetic probes have been used for the $\eta$
production processes, essentially off the nucleon.
The data for $\pi N\to \eta N$ come mainly from measurements performed in
70s~\cite{Deinet1969,Richards1970,Debenham1975,Brown1979}
and suffer from some inconsistencies~\cite{Clajus1992}, except a recent
experiment performed at BNL, using the Crystal Ball spectrometer~\cite{Morrison1999,Prakhov2005}.
The published CB data~\cite{Prakhov2005} are high quality measurements, though limited to the close
to threshold kinematics.
For the photoproduction process, a healthy amount of data has been published in recent years
for both differential cross section~\cite{Dugger2002,Crede2005,Nakabayashi2006,Bartalini2007} and
polarized beam asymmetry~\cite{Elsner2007,Bartalini2007}.
Recently CBELSA/TAPS~\cite{Crede2009} and CLAS~\cite{Williams2009} Collaborations released new data with
higher precision and higher energies~\cite{Williams2009}.

In spite of the unbalanced data base between $\gamma N$ and $\pi N$ initiated reactions,
recent theoretical efforts~\cite{Feuster1998,Feuster1999,Penner2002,Penner2002d,
Anisovich2005,Sarantsev2005,Batinic1997,Ceci2006,Arndt2007,Vrana2000,Matsuyama2007,Chen2003,Chen2007a,Gasparyan2003,Julia-Diaz2006,Julia-Diaz2007,Durand2008,Durand2009,Chiang2003,Chiang2004}
are focusing, via coupled-channels approaches, on both
families of processes, taking into account a significant number of intermediate and/or
final states, such as
$\pi N,~\eta N,~\rho N,~\sigma N,~\pi \Delta,~K \Lambda,~K \Sigma$.
Those works are all based on effective Lagrangian approaches, where meson-baryon degrees of freedom are implemented.

Investigations based on subnucleonic degrees of freedom, via chiral constituent quark models
($\chi$CQM) have been developed and successfully applied to the interpretation of photoproduction
of pseudoscalar mesons on the proton, namely, $\gamma p \to \pi N$~\cite{Zhao2002},
$\eta p$~\cite{Saghai2001,He2008,He2008a}, $K^+ \Lambda$ ~\cite{Julia-Diaz2006}, as well as the
process $\pi^- p \to \eta n$~\cite{He2009,Zhong2007}.
Finally, in a recent paper a formalism in the $\chi$CQM approach~\cite{He2008a} was
extended to the $\pi^-p\rightarrow \eta n$ reaction and allowed performing a combined
analysis of both channels~\cite{He2009}.

The previous studies within the $\chi$CQM approach focus on the total-energy ($W$) region below
2 GeV, where the best known $N^*$s lie.
In those formalisms contribution from  $N^*$s of $n>2$ shells in harmonic oscillator basis,
relevant to the higher energy region, are treated as degenerate, that is, simulated by a
single resonance.
Recent data extending the phase space up to $W \approx$2.8 GeV renders the degenerate
approximation less reliable. One solution to that issue is to use
the duality assumption~\cite{Collins} and to complement the $\chi$CQM with the {\it t}-channel contributions, which
mimic~\cite{Saghai1996} the left-out higher mass resonances ($M >$ 2 GeV).
Formalisms using only Regge trajectories or associating the effective Lagrangian for
{\it s-}channel with a Reggeized model
for {\it t-}channel have proven to be successful in describing data above
$W \approx$2 GeV for pseudoscalar mesons:
$\pi$~\cite{Guidal1997a,Sibirtsev2007a,Sibirtsev2009,Sibirtsev2009a},
$\eta$~\cite{Chiang2003} and
kaon~\cite{Guidal1997a,Guidal2003,Corthals2006,Corthals2007,Vancraeyveld2009a,Vancraeyveld2010}
photoproduction reactions.
Besides, the MAID Group~\cite{Chiang2003} showed that a Reggeized model is more appropriate than the
{\it t-}channel exchanges described by the $\rho$ and $\omega$ poles.

So, in this work we adopt the Reggeized model to complement our $\chi$CQM
formalism~\cite{He2008a,He2009} and study the $\eta$ production, especially the very recent
data released by CBELSA/TAPS~\cite{Crede2009} and CLAS~\cite{Williams2009} Collaborations.

The paper is organized as follows.
In Section~\ref{Sec:Theo}, the theoretical frame of our $\chi CQM$ approach, for
both $\gamma p \to \eta p$ and $\pi^- p \to \eta n$, and the Reggeized
model are briefly presented.
The fitting procedure and numerical results for differential cross section,
polarized beam asymmetry, helicity amplitudes, and partial decay widths are
reported and discussed in section~\ref{Sec:Res}.
Summary and conclusion are reported in section~\ref{Sec:Conclu}.

\section{Theoretical Frame}\label{Sec:Theo}

Chiral quark model approaches, are based on the low energy QCD
Lagrangian~\cite{Manohar1984,Li1997}

\begin{eqnarray} \label{lg}
\mathcal{L}=\bar{\psi}[\gamma_{\mu}(i\partial^{\mu}+V^{\mu}+\gamma_5A^{\mu})-m]\psi
+\cdot\cdot\cdot,
\end{eqnarray}
where vector ($V^{\mu}$) and axial ($A^{\mu}$) currents read,
\begin{eqnarray}
V^{\mu} =
 \frac{1}{2}(\xi\partial^{\mu}\xi^{\dag}+\xi^{\dag}\partial^{\mu}\xi)~,~A^{\mu}=
 \frac{1}{2i}(\xi\partial^{\mu}\xi^{\dag}-\xi^{\dag}\partial^{\mu}\xi),
\end{eqnarray}
with $ \xi=\exp{(i \phi_m/f_m)}$ and $f_m$ the meson decay constant.
$\psi$ and $\phi_m$ are the quark and meson fields, respectively.

The amplitudes for {\it s}-channel resonances can then be written as,

\begin{equation}\label{42}
{\cal M}_{N^*}=\frac {2M_{N^*}}{s-M_{N^*}^2-iM_{N^*}\Gamma({\bf q})}e^{-\frac {{\bf
k}^2+{\bf q}^2}{6\alpha_{ho}^2}} {\cal O}_{N^*},
\end{equation}
where $\sqrt {s} \equiv W$ is the total
energy of the system, and ${\cal O}_{N^*}$ is determined by the
structure of each resonance. $\Gamma({\bf q})$ in Eq. (\ref{42}) is
the total width of the resonance, and a function of the final state
momentum ${\bf q}$.

The transition amplitude for the $n^{th}$ harmonic-oscillator shell is
\begin{eqnarray}\label{On}
{\cal O}_{n}={\cal O}_n^2 +{\cal O}_n^3.
\end{eqnarray}

The first (second) term represents the process in which the incoming
photon and outgoing meson are absorbed and emitted by the
same (different) quark~\cite{Li1997,Zhong2007}.

We use the standard multipole expansion of the CGLN amplitudes~\cite{Chew1957}
to obtain the partial wave amplitudes for resonance $f_{2I, 2l\pm1}$.
Then the transition amplitudes for pseudoscalar meson production through photon and
meson induced reactions take, respectively, the following forms:
\begin{eqnarray}
\label{63} {\cal O}^\gamma_{N^*}&=&if_{1l\pm}  {\bf \sigma} \cdot {\bf
\epsilon}+ f_{2l\pm} {\bf \sigma} \cdot {\bf \hat{q}} {\bf \sigma}
\cdot ({\bf \hat{k}} \times {\bf \epsilon})+ if_{3l \pm} {\bf
\sigma} \cdot {\bf \hat{k}} {\bf \hat{q}} \cdot {\bf \epsilon} +
if_{4l\pm}  {\bf \sigma} \cdot {\bf \hat{q}}{\bf \epsilon}\cdot {\bf
\hat{q}},\\
{\cal O}^m_{N^*}&=&f_{1l\pm}+{\bf
\sigma}\cdot\hat{\bf q}{\bf \sigma}\cdot\hat{\bf k}f_{2l\pm}.
\end{eqnarray}

We can relate the helicity amplitude for a given resonance with the
multipole coefficient as in the case of photoproduction process~\cite{He2009}
\begin{eqnarray}
	f_{l\pm}^{N^*}&=& \mp A^{N^*}_{l\pm}=\frac{1}{2}\epsilon \left(
	\frac{\Gamma_{\pi N} \Gamma_{\eta N}}{kq} \right)^{1/2}C^{N^*}_{\pi N}C^{N^*}_{\eta N} \;
= \frac{1}{2\pi(2J+1) }
\left( \frac{ E_{N_i}E_{N_j}}{
M^2_{N^*}} \right)^{1/2}A^{\pi N}_{1/2}A^{\eta N}_{1/2},
\end{eqnarray}
where the decay width is given by
\begin{eqnarray}
\Gamma_m=\frac{1}{(2J+1) }\frac{| q|E_N}{\pi M_{N^*}} {\left | \frac {A^m_{1/2}} {C^{N^*}_{m N}}   \right |}^2,
\end{eqnarray}
and the Clebsch-Gordan coefficients are
\begin{eqnarray}
C^{N^*}_{m N}=\langle I^{N^*}M^{N^*}|I^{m}M^{m}I^NM^N\rangle,
\end{eqnarray}
with $m \equiv \pi,~\eta$.

In our approach, the photoexcitation helicity amplitudes $A_\lambda^\gamma$, as well as
the strong decay
amplitudes $A^m_\nu$, are related to the matrix elements of the interaction Hamiltonian~\cite{Copley1969}
as following:
\begin{eqnarray}
A^\gamma _\lambda  &=&\sqrt{\frac{2\pi}{k}}\langle
{N^*};J\lambda|H_{e}|N;\frac{1}{2}\lambda-1\rangle\,, \label{Agam}\\
A^m_\nu&=&\langle N;\frac{1}{2}\nu|H_{m}|{N^*};J\nu\rangle. \label{Ames}
\end{eqnarray}

For the {\it s}-channel, Eqs.~(\ref{Agam}) and (\ref{Ames}) are used to determine contributions from
$n \leq$2 shells for $N^*$s with $M \lesssim$2 GeV.
The {\it u}-channel is calculated as degenerate as before~\cite{Saghai2001,He2008a,He2009,Li1997}.
For the {\it t}-channel contributions, we start with Feynman diagrams for
vector mesons  $\rho^0$ and $\omega$ exchanges amplitudes.
The effective Lagrangian for the vector meson exchange vertices are

\begin{eqnarray}
 {\mathcal{L}}_{\gamma \eta V} & = & \frac{e \lambda_V}{m_{\eta}}\,
 \varepsilon_{\mu \nu \rho \sigma}\,(\partial^{\mu} A^{\nu}_\gamma)\,\phi_{\eta}\,
 (\partial^{\rho} V^{\sigma})\,, \label{eq:LgeV}
 \\ [1ex]%
 {\mathcal{L}}_{V qq} & = &g_V {\bar{\psi}} \left(  \gamma_{\mu} +
 \frac{\kappa_Vqq}{2 m_q}\, \sigma_{\mu \nu}
 \partial^{\nu} \right) V^{\mu}_V \psi \,, \label{eq:LVNN}
\end{eqnarray}
where $A^\nu _\gamma$ and $V^\mu _V$ are photon and exchanged vector mesons, respectively.
Here, in line with Ref.~\cite{Zhao2002} we introduced $\kappa_Vqq$ for the constituent quark.
The electromagnetic couplings of the vector mesons $\lambda_V$ are determined from the radiative
decay widths $\Gamma_{V\rightarrow\eta\gamma}$ as MAID group~\cite{Chiang2003},
that is, 0.81 for $\rho$ and $0.291$ for $\omega$.
The values for the strong coupling constants $g_{Vqq}$  and $\kappa_{Vqq}$ are treated as free parameters.

With the Lagrangians in Eqs. (\ref{eq:LgeV}) and (\ref{eq:LVNN}), the amplitudes for {\it t}-channel in chiral quark model can be easily obtained
as in Ref.~\cite{Zhao2002}
\begin{eqnarray}
\label{vme}
{\cal M}_V&=&e\frac{\lambda_V g_{Vqq}
e^{-({\bf k}-{\bf q})^2/6\alpha_{ho}^2}}
{m_\pi (t-m^2_V)}
\left\{ g_t\left[1+\frac{\omega_m}{E_f+M_f}
+\frac{\omega_\gamma}{E_i+M_i}\right.\right.\nonumber\\
&&\left. +\frac{\kappa_{Vqq}}{2m_q}\left(\frac{m^2_\pi}{E_f+M_f}
-\left(\frac{1}{E_f+M_f}+\frac{1}{E_i+M_i}\right)k\cdot q\right)\right]
{\bf q}\cdot({\bf k}\times{\bm \epsilon}_\gamma)\nonumber\\
&&+g_A
\left[\frac{\omega_\gamma {\bf q}^2}{E_f+M_f} +
\frac{\omega_m {\bf k}^2}{E_i+M_i}
-\left(\frac{\omega_\gamma}{E_i+M_i}+\frac{\omega_m}{E_f+M_f}\right)
{\bf q}\cdot{\bf k} \right.\nonumber\\
&&+\frac{\kappa_{Vqq}}{2m_q}\left(\omega_m {\bf k}^2 +\omega_\gamma {\bf q}^2
+\frac{\omega_\gamma\omega_m}{E_f+M_f}{\bf q}^2
+\frac{(\omega_\gamma\omega_m-m^2_\pi)}{E_i+M_i}{\bf k}^2\right.\nonumber\\
&&\left.\left. -\left(\omega_\gamma+\omega_m
+\frac{\omega^2_m}{E_f+M_f}+\frac{\omega^2_\gamma}{E_i+M_i}
-\frac{k\cdot q}{E_i+M_i}+\frac{{\bf q}\cdot{\bf k}}{E_f+M_f}\right)
{\bf q}\cdot {\bf k}\right)\right]i{\bm \sigma}\cdot{\bm \epsilon}_\gamma\nonumber\\
&&+g_A\left[\frac{\omega_m}{E_f+M_f}
+\frac{\kappa_{Vqq}}{2m_q}\left(\omega_m+\frac{\omega^2_m}{E_f+M_f}
-\frac{k\cdot q}{E_f+M_f}+\frac{{\bf k}\cdot{\bf q}}{E_i+M_i}\right)\right]
i{\bm \sigma}\cdot{\bf k}{\bf q}\cdot{\bm \epsilon}_\gamma \nonumber\\
&&\left. -g_A\left[\frac{\omega_\gamma}{E_f+M_f}
+\frac{\kappa_{Vqq}}{2m_q}\left(\omega_\gamma+\frac{{\bf k}^2}{E_i+M_i}
+\frac{\omega_\gamma\omega_m}{E_f+M_f}\right)\right]i{\bm \sigma}\cdot{\bf q}
{\bf q}\cdot{\bm \epsilon}_\gamma
\right\} \ ,
\end{eqnarray}
where $k\cdot q=\omega_\gamma\omega_m-{\bf k}\cdot {\bf q}$ is the
four-momentum product. As in Ref.~\cite{Li1997}, a Lorentz boost is
adopted here, which was introduced to inlude the
relativistic effect in the calcualtion of the proton form
factor \cite{Licht1970}. The exponential term $e^{-({\bf k}-{\bf
q})^2/6\alpha_{ho}^2}$ comes from the nucleon wave function, which
plays the role of a form factor. However, the MAID
Group~\cite{Chiang2003} study showed that the form factor is not
required for the Reggeized model.  
Hence, we remove the exponential
term in our calculations (see \ref{Sec:gamma}).  $g_A$ is the axial vector coupling and is
defined in the quark model as $\langle N_f | \sum_j{\hat I}^v_j {\bm
\sigma}_j|N_i \rangle \equiv g_A\langle N_f | {\bm \sigma}|N_i \rangle
,$ where ${\hat I}^v_j$ is the isospin operator for the exchanged
vector meson.  The factor $g_t$ comes from the isospin space,
$g_t\equiv \langle N_f | \sum_j{\hat I}^v_j |N_i \rangle$.

In the Reggeized model the main change is substituting the meson exchange poles by the
Regge propagator:
\begin{eqnarray}
	\frac{1}{t-m_V^2} \to {\cal P}^V_{Regge}= \left(\frac{s}{s_0} \right)^
	{\alpha_V(t)-1}\frac{\pi\alpha'_V}{\sin \left[\pi\alpha_V(t) \right]}
	\frac{ {\cal S}_V+e^{-i\pi\alpha_V(t) }}{2}
	\frac{1}{\Gamma \left(\alpha_V(t)\right)},
\end{eqnarray}
where $s_0$=1 GeV$^2$ is the reference mass scale and ${\cal S}_V=\pm 1$ is the trajectory's signature.
The gamma function $\Gamma(\alpha_V(t))$ suppresses poles of the propagator in the unphysical region.
The vector-meson Regge trajectory is taken in the following linear form
\begin{eqnarray}
 \alpha_V (t)=\alpha_V^\circ+\alpha_V^\prime t,
\end{eqnarray}
with $t$ the Mandelstam variable, and read for $\rho$ and $\omega$, respectively, as
\begin{eqnarray}
 \alpha_{\rho}(t)&=&0.55+0.8 t,\\
 \alpha_{\omega}(t)&=&0.44+0.9 t.
\end{eqnarray}

\section{Results and discussion}\label{Sec:Res}
\subsection{Fitting procedure}\label{Sec:Fit}
Using the CERN MINUIT code, we have fitted simultaneously the following data sets and PDG values:

\begin{itemize}
\item {\bf Spectrum of known resonances:}
    \subitem {\it Known resonances:}
 We use as input the PDG values~\cite{Amsler2008} for masses and widths, for which the
uncertainties are handled as in Ref.~\cite{He2008a}.
The 12 known nucleon resonances, with $M~\lesssim$~2 GeV, considered in this work are:

{\boldmath$ n$}{\bf =1:} $S_{11}(1535)$, $S_{11}(1650)$,
$D_{13}(1520)$, $D_{13}(1700)$, and $D_{15}(1675)$;

{\boldmath$ n$}{\bf =2:} $P_{11}(1440)$, $P_{11}(1710)$,
 $P_{13}(1720)$, $P_{13}(1900)$,
$F_{15}(1680)$, $F_{15}(2000)$, and $F_{17}(1990)$.

Besides the above isospin-1/2 resonances, we fit also the mass of the
$\Delta$(1232) resonance. However, spin-3/2 resonances do not
intervene in the $\eta$ photoproduction.
\item {\bf  Observables for $\gamma p\rightarrow \eta p$:}

\subitem {\it Differential cross-section:} Data base includes 3349 data points
for 1.5 $\lesssim W \lesssim$ 2.8 GeV, coming from the following labs:
MAMI~\cite{Krusche1995}, CLAS~\cite{Dugger2002}, ELSA~\cite{Crede2005},
LNS~\cite{Nakabayashi2006}, and GRAAL~\cite{Bartalini2007}. Only statistical
uncertainties are used.  For the most recent data from
CBELSA/TAPS~\cite{Crede2009} and CLAS~\cite{Williams2009} Collaborations,
both statistical and systematic uncertainties are considered to avoid too strong constraints
due to their very small statistical uncertainties.
\subitem {\it Polarization observables:}
For polarized beam asymmetry $\Sigma$, 184 data points for 1.5 $\lesssim W
\le$ 1.9 GeV from GRAAL~\cite{Bartalini2007} and ELSA~\cite{Elsner2007}
are used with statistical uncertainties.

The target asymmetry ($T$) data~\cite{Bock1998} are not included in our data base.
Actually, those 50 data points bear too large uncertainties to put significant
constraints on the parameters~\cite{He2008a}.
\item {\bf  Observables for $\pi^-p \to \eta n$:}
In line with Ref.~\cite{Durand2008}, the used data base includes 354 differential cross sections,
for 1.5 $\lesssim~W~\le$ 2.0 GeV, coming from: Deinet~\cite{Deinet1969},
Richards~\cite{Richards1970}, Debenham~\cite{Debenham1975}, Brown~\cite{Brown1979},
Prakhov~\cite{Prakhov2005}.
Uncertainties are treated as in Ref.~\cite{Durand2008}
\end{itemize}
In summary, 3887 experimental values are fitted.
To do so, we have a total of 19 free parameters, not all of them adjusted on all the data sets,
as explained below.

In Table~\ref{Tab: Para} we summarize the list of adjustable parameters and their extracted values.
Note that the reported uncertainties are those produced by the MINUIT code and should be considered
as lower limits.
%
%
\begin{table*}[ht!]
\caption{{\footnotesize Adjustable parameters with their extracted values, where
$m_q$, $\alpha_{ho}$,  $\Omega$, $\Delta$, $M$, and $\Gamma$ are in
MeV.}}
\renewcommand\tabcolsep{0.5cm}
\begin{tabular}{llcc}  \hline\hline
	& Parameter                    &  Ref.~\cite{He2009} & Present work      \\ \hline
  & $m_q$                        &  312                & 310 $\pm$ 5   \\
  & $\alpha_{ho}$                &  348                & 309 $\pm$ 2   \\
  & $\alpha_s$                   &  1.96               &  1.60 $\pm$ 0.02  \\
  & $\Omega$                     &  437                & 421 $\pm$ 4   \\
  & $\Delta$                     &  460                & 460 $\pm$ 1   \\ \hline
  & $g_{\eta NN}$                &  0.376              & 0.276 $\pm$ 0.005 \\ \hline
$P_{13}(1720)$:
  & $C^\gamma_{P_{13}(1720)}$    &  0.37               & 0.22 $\pm$ 0.01   \\
  & $C^\pi_{P_{13}(1720)}$       & -0.89               & -0.85 $\pm$ 0.03  \\ \hline
New $S_{11}$: &$M^\gamma$        & 1715                & 1700 $\pm$ 1  \\
              &$\Gamma^\gamma$   & 207                 & 473 $\pm$ 10  \\
              &$C^\gamma_{N^*}$  & 0.51                & 1.18 $\pm$ 0.03   \\ \hline
$N(1535)$:    &$M^\gamma$        & $--$                & 1532 $\pm$ 1  \\
              &$\Gamma^\gamma$   & $--$                & 140 $\pm$ 1   \\\hline
$u$-channel:
  & $C^\gamma_{u}$               & $--$                & 0.71 $\pm$ 0.03   \\
  & $C^\pi_{u}$                  & $--$                & 1.39 $\pm$ 0.05   \\ \hline
{\it t}-channel   &$g_{\rho qq}$     & $--$            & 1.90 $\pm$ 0.22 \\
	      &$\kappa_{\rho qq}$     & $--$                 &-0.20 $\pm$ 0.01 \\
	      &$g_{\omega qq}$   & $--$                      & 4.88 $\pm$ 0.16 \\
	      &$\kappa_{\omega qq}$   & $--$                 & -0.26 $\pm$ 0.02 \\\hline
 \hline\end{tabular} \label{Tab: Para}
\end{table*}

Two of the parameters, namely, the non-strange quarks average mass ($m_q$) and
the harmonic oscillator strength ($\alpha_{ho}$) are involved in fitting both mass
spectrum and $\eta$-production data.
The QCD coupling constant ($\alpha_s$) and the confinement constants ($\Omega$ and $\Delta$),
intervene only  in fitting the $\eta$-production data {\it via} the configuration mixing mechanism.
In Table~\ref{Tab: Para} the extracted values within the present work are given and compared
to those reported in our previous paper~\cite{He2009} where a single higher mass resonance was
considered, corresponding to a degenerate treatment of $n>$2 shells nucleon resonances,
instead of the {\it t}-channel mechanism.
The quark mass and the harmonic oscillator strength ($\alpha_{ho}$) are close to the values in the previous fitting~\cite{He2009} while the QCD coupling constant ($\alpha_s$) gets decreased by
about 17\%.
For the other parameters, the extracted values come out close to those used by
Isgur-Karl~\cite{Isgur1979} and Capstick-Roberts~\cite{Capstick2000}:
$E_0$~=~1150 MeV, $\Omega \approx$ 440 MeV, and $\Delta \approx$ 440 MeV.

The remaining parameters are involved in the fitting of $\eta$-production data.
With respect to the $\eta$-nucleon coupling constant $g_{\eta NN}$, our
result favors a rather small coupling around $g_{\eta NN} = 0.3$, which is
compatible with those deduced from fitting only the $\eta$
photoproduction~\cite{Li1998,Saghai2001}.  Comparable values for that coupling
are also reported in Refs.~\cite{Tiator1994,Kirchbach1996,Zhu2000,Stoks1999}.

As already mentioned, the $S_{11}(1535)$ is the dominant resonance in the $\eta$ production processes
up to $W \lesssim$1.7 GeV.
Hence we treat its Breit-Wigner mass and width as free parameters.
Both extracted values come out within the PDG ranges: $M=1535 \pm 10$ MeV and $\Gamma=150 \pm 25$ MeV.

Two strength parameters are introduced for the resonance ${P_{13}(1720)}$, in line with
Ref.~\cite{He2008a}, and treated as adjustable in order to avoid its
otherwise too large contribution resulting from direct calculation.
The value of that parameter extracted from the photoproduction reaction, is close to that
obtained~\cite{He2008a} by fitting data below $W \lesssim$ 2 GeV.

In our previous works~\cite{He2008a,He2009}, three new resonances, $S_{11}$, $D_{13}$, and $D_{15}$,
were introduced, in line with findings by several authors
~\cite{Saghai2001,Sarantsev2005,Li1996,Batinic1997,Giannini2001,Ablikim2006,
Fang2006,Chen2003,Chiang2003,Tryasuchev2004,Mart2004,Kelkar1997,Julia-Diaz2006},
with extracted masses roughly between 1.7 and 2.1 GeV.
The new $D_{13}$ state was found to be negligible in the $\eta$ photoproduction~\cite{He2008a,He2009}
and, hence, is not  considered here.
The contribution from $D_{15}$ ($M >$2 GeV) might overlap with the {\it t}-channel contributions
in this work.
Hence we have removed it to avoid double-counting problem.
The third $S_{11}$ is still kept in this work  and its extracted
mass comes out close to the previous works~\cite{Saghai2001,He2008a,He2009} while the width is
about 470 MeV, which is  larger than the value in Ref.~\cite{He2008a}. Such a large width might be
an indication of additional $S_{11}$ resonance(s), as reported in Ref.~\cite{Chen2003}.
Finally, for the process $\pi^-p \to \eta n$, we deal only with the known resonances,
as in Refs.~\cite{Durand2008,He2009}.

Now we discuss the {\it u}- and {\it t}-channels treatments.

In chiral quark model approaches, the $u$-channel contribution is
handled as degenerate~\cite{Li1997}.
In the work of MAID group~\cite{Chiang2003}, it was suggested that the $u$-channel contributions
may be important in reproducing the behavior of differential cross section at extreme backward angles.
Hence we introduce a unique global adjustable parameter for this channel.
Numerical results show about 30\% deviations from unity, expected within exact $SU(6) \otimes O(3)$
symmetry.

Below, we give the expressions relating the meson-nucleon-nucleon couplings to those at the quark level:
\begin{eqnarray}
g^V_t g_{Vqq}&=&g_{VNN},  \label{Vt}\\
g^V_A\frac{g_{Vqq}}{m_q}(1+\kappa_{Vqq})&=&\frac{g_{VNN}}{m_N}(1+\kappa_{VNN})
\label{VA} .
\end{eqnarray}
Values for $g_{Vqq}$ and $\kappa_{Vqq}$, with $V \equiv \rho,~\omega$, have been extracted by fitting
the photoproduction data, as reported in Table \ref{Tab: Para}.
For the {\it t}-channel small values of $\kappa_{Vqq}$ are found as expected.
For the remaining parameters we use the quark model values~\cite{Zhao2002}:
$g^\omega_A=1$, $g^\omega_t=3$,
$g^{\rho^\circ}_A=5/3$ and $g^{\rho^\circ}_t=1$.
Those values are comparable with the values reported by other
authors~\cite{Drechsel1999,Chiang2003,Davidson1991,Dumbrajs1983,Gasparyan2003},
within the uncertainties on extracted values for $g^A$ and $g^t$ in constituent quark model.

Using our extracted values for $Vqq$ vertices couplings (Table~\ref{Tab: Para}, last four rows) and
Eqs.~(\ref{Vt}) and (\ref{VA}) lead, for $VNN$ vertices, to results reported in Table~\ref{Tab: cc} and
compared with results from three other works. Our values for $\rho$ are smaller than those extracted by the
MAID analysis~\cite{Chiang2003} and also coming from Bonn~\cite{Machleidt1987}
and Nijmegen~\cite{Rijken2006} potentials. Note that results from these latter potentials differ between
themselves by 30\% to 50\%. For $\omega$ case our value for $g_{\omega NN}$ comes out larger than
the MAID analysis~\cite{Chiang2003}, but stands in-between those produced by nucleon-nucleon potentials.
The $\kappa_{\omega NN}$ deviates significantly from vanishing values of potentials.
Detailed discussion on the $VNN$ couplings values can be found in Refs.~\cite{Penner2002,Penner2002d}.
%
%
\begin{table*}[ht!]
\caption{{\footnotesize Vector-meson nucleon-nucleon couplings.}}
\renewcommand\tabcolsep{0.5cm}
\begin{tabular}{lcccc}  \hline\hline
Ref.	           & $g_{\rho NN}$ & $\kappa_{\rho NN}$ & $g_{\omega NN}$ & $\kappa_{\omega NN}$ \\ \hline
Present work                        & 1.90 & 3.0 & 14.6 & -0.25 \\
MAID~\cite{Chiang2003}              & 2.4  & 3.7 & 9.0  &  0    \\
Bonn potential~\cite{Machleidt1987} & 3.34 & 6.1 & 15.8 &  0    \\
Nijmegen potential~\cite{Rijken2006}& 2.76 & 4.2 & 11.1 & 0.02  \\\hline
 \hline\end{tabular} \label{Tab: cc}
\end{table*}

At this stage, having presented various parameters of our model, either fitted or taken from
literature, we proceed to rather detailed discussion, per data set, of the quality of our fit.

In Table \ref{Tab: chi2}, rows 2 to 7 give the  total $\chi^2$ (column 4) and the $\chi^2$
per data point (column 6) for $\gamma p \to \eta p$ differential cross sections.
The $\chi^2_{dp}$ comes out smaller than 2 for data published between years 1995 and 2007,
by collaborations from MAMI~\cite{Krusche1995} (MAMI95), LNS~\cite{Nakabayashi2006} (LNS06)
and GRAAL~\cite{Bartalini2007} (GRAAL07). Those data span a total energy range going from
threshold to $W \lesssim$1.9 GeV.
%
\begin{table*}[hb!]
\caption{{\footnotesize $\chi^2$ for the $\gamma p \to \eta p$ differential cross section (rows 2 to 8)
and polarized beam asymmetry (rows 9 and 10); differential cross section of the reaction
$\pi^- p \to \eta n$ (rows 11 and 15); and mass spectrum of isospin 1/2 baryon resonances (row 16).}}
\renewcommand\tabcolsep{0.35cm}
\begin{tabular}{l|lrrrc}  \hline\hline
Observable & Collaboration/author  & W (GeV)           & $\sum\chi^2$ & $N_{dp}$ &    $\sum\chi^2_{dp}$\\ \hline
$\frac{d\sigma}{d\Omega}~(\gamma p \to \eta p)$
           &MAMI94~\cite{Krusche1995} &   1.49 - 1.54 &  183.73 & 100 & 1.84 \\
           &CLAS02~\cite{Dugger2002} &   1.53 - 2.12 &  929.49 & 190 & 4.90 \\
           &CLAS09~\cite{Williams2009} &   1.68 - 2.80 & 2595.48 &1081 & 2.40 \\
           &ELSA05~\cite{Crede2005} &   1.53 - 2.51 & 1250.66 & 631 & 1.98 \\
           &ELSA09~\cite{Crede2009} &   1.59 - 2.37 & 2028.88 & 680 & 2.98 \\
           &LNS06~\cite{Nakabayashi2006} &    1.49 - 1.74 &  313.16 & 180 & 1.74 \\
           &GRAAL07~\cite{Bartalini2007} &    1.49 - 1.91 &  629.13 & 487 & 1.29 \\\hline
$\Sigma~(\gamma p \to \eta p)$
           &ELSA07~\cite{Elsner2007} &    1.57 - 1.84 &  42.21  & 34  &1.24  \\
           &GRAAL07~\cite{Bartalini2007} &    1.50 - 1.91 & 883.60  & 150 &5.89  \\\hline
$\frac{d\sigma}{d\Omega}~(\pi^-p\rightarrow \eta n)$
           &Prakhov~{\it et al.}~\cite{Prakhov2005}&    1.49 - 1.52 &   39.45 & 84 &0.47 \\
           &Deiinet~{\it et al.}~\cite{Deinet1969} &     1.51 - 1.70 &  127.01 & 80 &1.59 \\
           &Richards~{\it et al.}~\cite{Richards1970} &     1.51 - 1.90 &  122.31 & 64 &1.91 \\
           &Debenham~{\it et al.}~\cite{Debenham1975} &     1.49 - 1.67 &   16.71 & 24 &0.70 \\\
           &Brown~{\it et al.}~\cite{Brown1979} &     1.51 - 2.45 &  159.82 &102 &1.57 \\\hline
$N^*$ Spectrum & PDG~\cite{Amsler2008}  &             & 51.2    & 15 &3.42 \\\hline
Total      &      &                 & 9372.2  &3902&2.40 \\
\hline \hline\end{tabular} \label{Tab: chi2}

\end{table*}

The best reduced $\chi^2$ is obtained for GRAAL07 measurements,
with close to 500 data points. The worse $\chi^2_{dp}$ is given by the first results
published~\cite{Dugger2002} in 2002 by the CLAS Collaboration (CLAS02), which show discrepancies
with the most recent results~\cite{Crede2009,Williams2009}, including those by the same
Collaboration (CLAS09),
which extends the higher limit in energy from $W \approx$ 2.1 to 2.8 GeV and embodies almost
6 times more data points. Measurements performed at ELSA show different trends, namely,
the early work~\cite{Crede2005} of that collaboration published in 2005 (ELSA05), goes
slightly higher in energy than their latest results~\cite{Crede2009}, but shows a smaller
$\chi^2_{dp}$ by about 30\%, with comparable number of data points. We will come back to these
considerations in the next Sec.
Results for polarized beam asymmetry are reported in Table \ref{Tab: chi2}, rows 8 and 9.
We get an excellent agreement with the ELSA measurements~\cite{Elsner2007} (ELSA07), but
a large $\chi^2_{dp}$ for GRAAL07 data~\cite{Bartalini2007}.

In summary, with respect to the photoproduction observables the $\chi^2_{dp}$ turns out to be
around 2.5, with significant discrepancies within various data sets and/or observables.

Finally, in rows 9 to 13 in Table \ref{Tab: chi2}, our results for the strong channel
are given. The overall $\chi^2_{dp}$ is 1.3, though it shows significant variations
according to the data set, but still it stays below 2.

We will come back to these considerations in the next Sec.

To end this section, we present our results for resonances spectrum and roles played by those
resonances in the reaction mechanisms of the processes considered in this work.

\begin{figure}[hb!]
\includegraphics[bb=75 50 360 340 ,scale=0.99]{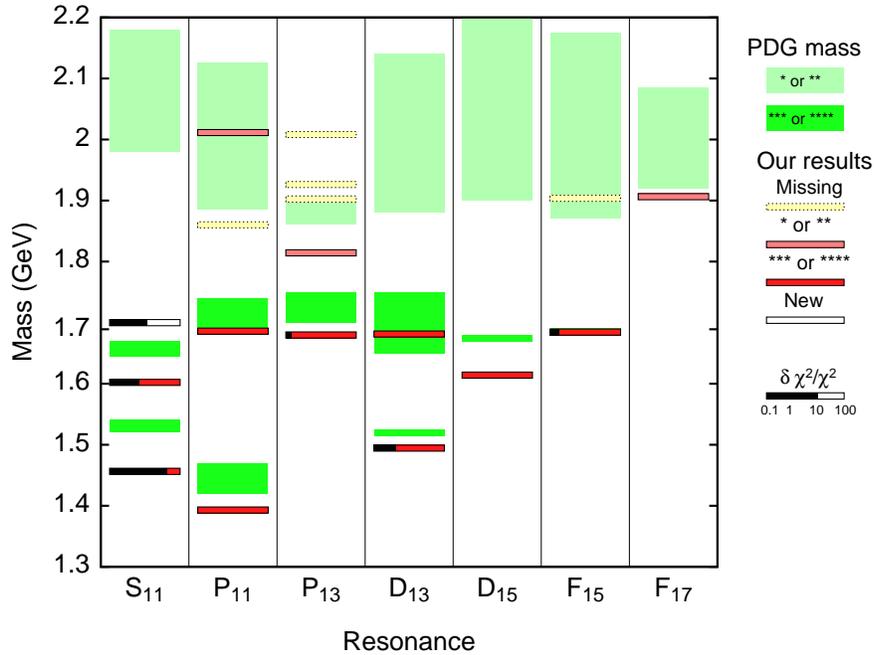}
\caption{(Color online) {\footnotesize The spectrum of baryon resonances from PDG~\cite{Amsler2008}
(green bands) and from the present work for known (red bans), missing (yellow bands),
and new (white band) resonances.
The black bands indicate the variations of $\chi^2$ after turning off the corresponding resonance
within our full model.}
\label{Fig:spcl}}
\end{figure}

In Fig.~\ref{Fig:spcl} are depicted mass values from PDG isospin 1/2
resonances with $M \lesssim$ 2 GeV, our results for the 12 known
$N^*$s listed in Sec.~\ref{Sec:Fit}, OGE generated missing
resonances~\cite{He2008a} ({\it $P_{11}$ (1899)}, {\it $P_{13}$
(1942)}, {\it $P_{13}$(1965)}, {\it $P_{13}$(2047)}, and {\it $F_{15}$
(1943)}) and the introduced third $S_{11}$.  The masses generated by
our formalism compare well enough with the PDG values in line with other CQM
approaches~\cite{Isgur1978a,Isgur1979}.  The same observation is valid
for missing resonances, as discussed in our previous
work~\cite{He2008a}.  The new $S_{11}$ has no counterpart within
known, neither missing resonances. 

In order to investigate the importance of the 18 resonances of our approach, we have switched off each of them one by one.
The five missing resonances show no significant effects in line with our
previous findings~\cite{He2008a}.
For the other $N^*$s, the black part in each bar (Fig.~\ref{Fig:spcl}) indicates the relative change
in $\chi^2$ with that specific resonance turned off.
Among the 12 known $N^*$s, the most significant ones, with decreasing importance, are:
$S_{11}(1535)$, $S_{11}(1650)$, $D_{13}(1520)$, $F_{15}(1680)$, $P_{13}(1900)$.
Finally, the new $S_{11}(1700)$ appears to be the second most important ingredient of our model.

In the following we  move to the observables for the $\gamma p \to \eta p$ and
$\pi^- p\rightarrow \eta n$ processes, and compare our results with the data.
To get better insights into the reaction mechanism, we also report on the contributions from
resonances which have significant effects on $\chi^2$ for both processes studied here.

\subsection{Observables for $\gamma p \to \eta p$}\label{Sec:gamma}

\begin{figure}[ht!]
\includegraphics[bb=35 150 560 640 ,scale=0.9]{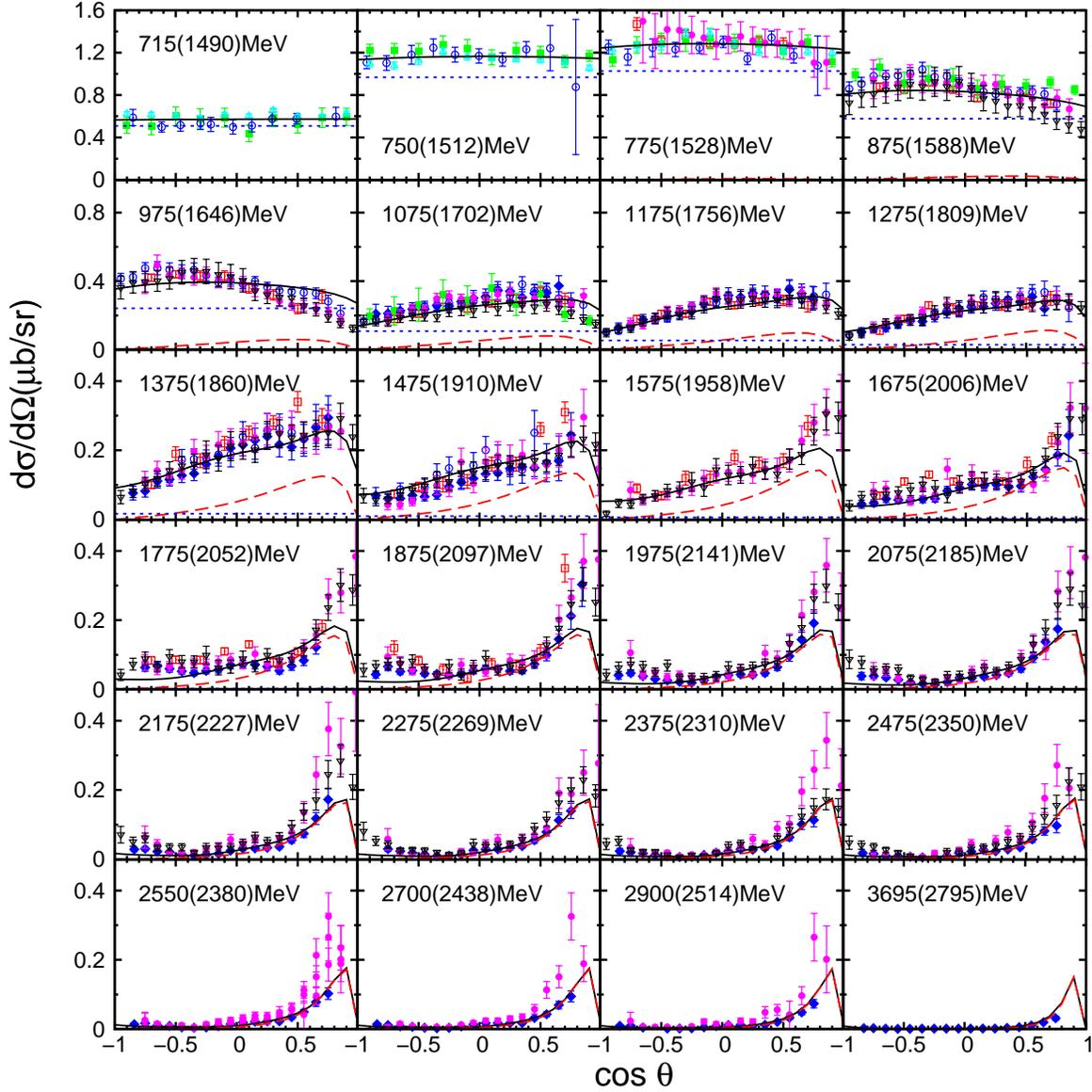}
\caption{(Color online) {\footnotesize Differential cross section for $\gamma p \to \eta p$ as
a function of $cos \theta_\eta$ for various values of photon energy in the lab frame.
The values in parenthesis are the corresponding total energy of the system $W$.
The curves are: full model (full),
$S_{11}(1535)$ (dotted), and {\it t}-channel (Dashed).
Data are from CLAS02 (open squares)~\cite{Dugger2002}, CLAS09 (filled
diamonds)~\cite{Williams2009}, LNS (filled squares)~\cite{Nakabayashi2006},
GRAAL (open circles)~\cite{Bartalini2007}, ELSA05 (filled circles)~\cite{Crede2005},
ELSA09 (down triangles)~\cite{Crede2009} and  MAMI (open triangles)~\cite{Krusche1995}.}
\label{Fig:gpep}}
\end{figure}

In Fig.~\ref{Fig:gpep} differential cross section results are depicted at twenty four energies going
from close to threshold, $E_{\gamma}^{lab}$=0.715 GeV ($W$=1.49 GeV) up to
$E_{\gamma}^{lab}$=3.70 GeV ($W$=2.80 GeV).
At each energy three curves are compared with the data: a) full model, b) contribution from solely
$S_{11}(1535)$, c) contributions from the Reggeized {\it t}-channel. Note that, while full model
embodies {\it u}-channel, its contribution is too small to be shown in the Figure.

Comparing our model (full curves) with various data shows that the general agreement is acceptable
and there is no anomalous behavior in the whole phase space.
Discrepancies within those data, in some cases with more than 2$\sigma$, make clear
problems that have to be faced in fitting such a data base, which is reflected in the $\chi^2_{dp}$=2.5.
The dotted curves show contributions due {\it only} to
$S_{11}(1535)$, which has a dominant role near to threshold and up to $W \approx$ 1.7 GeV,
where {\it t}-channel effects already start becoming visible. Note that the dashed curves correspond to
contributions {\it exclusively} from {\it t}-channel, without further minimizations.
From $W \approx$ 2 GeV on, this latter channel gains more and more importance with increasing energy
and completely dominates the model results above $W \approx$ 2.1 GeV.

In the range  2.2 $\lesssim W \lesssim$ 2.3 GeV the extreme angles are not very well reproduced, indicating
very likely that higher mass $N^*$s are needed and/or a more extended treatment of the {\it u}-channel
contributions is desirable.
In our work, {\it u}-channel contributions turn out to be very small in the whole energy region, with
a maximum contribution of roughly 8\%  around $W \approx$ 1.65 GeV and almost vanishing above
$W \approx$ 1.9 GeV.

\begin{figure}[ht!]
\includegraphics[bb=0 150 500 300 ,scale=0.95]{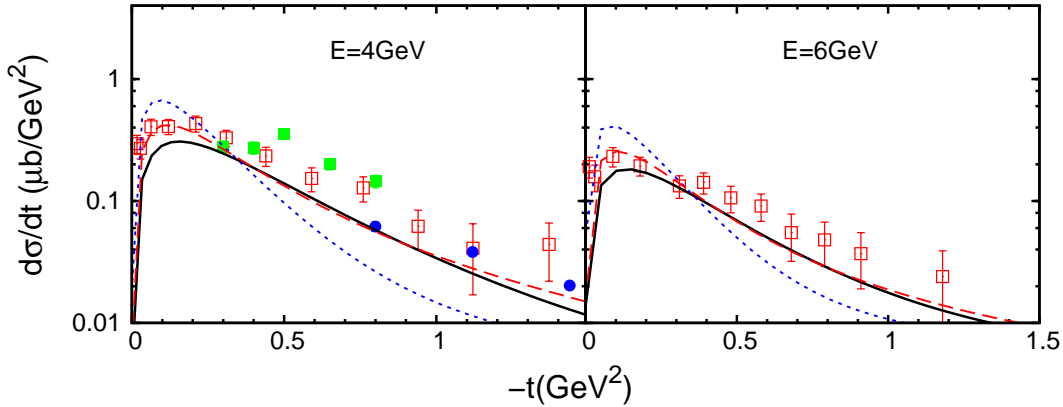}\
\caption{(Color online) {\footnotesize
Differential cross section for $\gamma p\to \eta p$ as a function of
$t$ at 4 and 6~GeV of photon energy in the lab frame. The solid
curves are our full model.
The dotted curves are obtained by keeping the expotential form factor (Eq. [14]), the dashed
curves are for $s_0=3$ GeV$^2$ with the expontential form factor kept.
Data are from Ref.~\cite{Braunschweig1970} (empty square), Ref.~\cite{Dewire1971} (filled
square) and Ref.~\cite{Williams2009} (fiiled circles).}
\label{Fig:gpep46}}
\end{figure}

Here, we would like to comment on our Regge treatment.
The canonical approach is to fix the Reggeized model parameters by fitting data at energies
above the resonance
region~\cite{Guidal1997a,Sibirtsev2007a,Corthals2006,Laget2005,Rodrigues2008,Sibirtsev2010}.
In this work, given that recent data cover both the resonance region and above, though not higher than 3.7 GeV, we have used only the recent data for consistency considerations. 
However, we report the predictions of our model at 4 and 6 GeV (Fig.~\ref{Fig:gpep46}, full curves)
and find reasonable agreement with the differential cross section data at high energies ~\cite{Braunschweig1970,Dewire1971}. 
From there, and the fact that data above $\approx$ 2 GeV show clearly the dominace of {\it t-}channel 
(Fig.~\ref{Fig:gpep}, dashed curves),  
we infer that the Regge approach parmeters extracted within the present wok, by fitting the recent data, are reliable enough.
In Fig.~\ref{Fig:gpep46} we show also the effects of the expotential
form factor in Eq.~\ref{vme},
that we removed. By adding that term to our full model, as expected we get a significant damping
effects (dotted curves), which can be compensated by increasing the mass scale $s_0$ from
1 to 3 GeV$^2$ (dashed curves).

\begin{figure}[ht!]
\includegraphics[bb=45 150 560 360 ,scale=0.95]{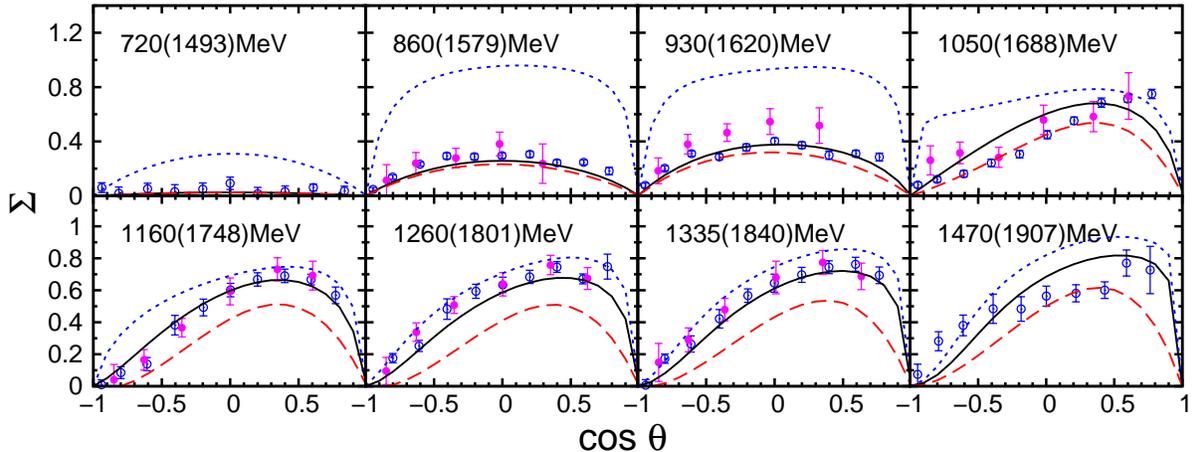}\
\caption{(Color online) {\footnotesize Same as Fig. \ref{Fig:gpep}, but for polarized
beam asymmetry for $\vec{\gamma} p\rightarrow \eta p$.
Data are from GRAAL (open circle)~\cite{Bartalini2007} and ELSA (filled circle)~\cite{Elsner2007}.}
\label{Fig: gpepBA}}
\end{figure}

To give a complete picture of the $\eta$ photoproduction, polarized beam asymmetries are presented
in Fig.~\ref{Fig: gpepBA}.
The full model (full curves) describes the data satisfactorily, except at $W=$1.688 and 1.907 GeV, which,
given the small uncertainties on GRAAL data, illustrates the large $\chi^2$ obtained for that data set.
Here, the two other curves have following ingredients: dotted curves come from the full model, with
$S_{11}(1535)$ switched off, and the dashed curves are obtained by turning off {\it t}-channel contributions. These two sets of results are also obtained without further minimizations.

As in the case of the differential cross section, the beam asymmetry is dominated by $S_{11}(1535)$
(dotted curves), up to $W \lesssim$ 1.7 GeV.
Above that energy, {\it t}-channel contributions become visible, with effects comparable
to differential cross section case in the same energy range.

\subsection{Observables for $\pi^-p\to\eta n$}\label{Sec:pion}
\begin{figure*}[ht!]
\includegraphics[bb=80 50 490 530 ,scale=0.95]{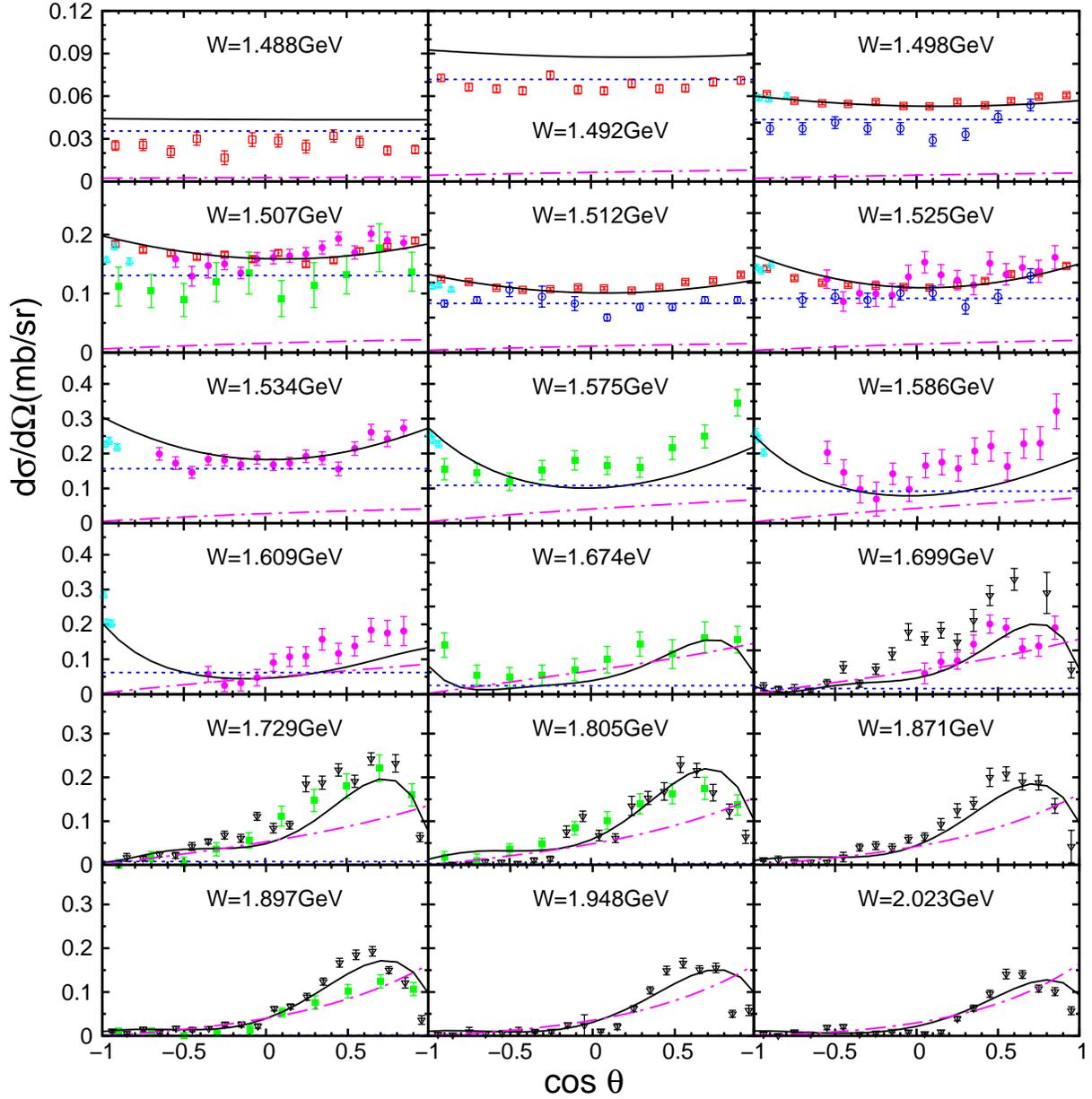}\
\caption{(Color online) {\footnotesize Differential cross section for $\pi^- p \to \eta n$ as a
function of $cos {\theta _\eta}$ for various values of $W$.
The curves are: full model (full curves),
$S_{11}(1535)$ (dotted), {\it u}-channel (dash-dotted).
Data are from  Prakhov {\it et al.} (open squares) \cite{Prakhov2005},
Richards {\it et al.} (filled squares) \cite{Richards1970},
Morrison {\it et al.} (open circles) \cite{Morrison1999},
Deinet {\it et al.} (filled circles) \cite{Deinet1969},
Debenham {\it et al.} (open triangles) \cite{Debenham1975},
Brown {\it et al.} (down triangles) \cite{Brown1979}.}
\label{Fig:ppen}}
\end{figure*}

In Fig.~\ref{Fig:ppen} we report our results for the full
model (full curves), contributions due to {\it only}
$S_{11}(1535)$ (dotted curves) and those {\it exclusively} from {\it u}-channel.

As already mentioned the data base for $\pi^-p \to \eta n$ suffers from serious inconsistencies, especially
with increasing energy. With this in mind, we limit ourselves to the $W \lesssim$ 2 GeV region.

The overall agreement between theory and heterogeneous data is acceptable.
The well known dominance of $S_{11}(1535)$ shows up at the lowest energies.
For $W\ge$1.6 GeV {\it u}-channel brings in significant contributions.
However, neither of these two ingredients reproduces the shape of the measured cross sections.
So here, the reaction mechanism again embodies destructive interferences among $N^*$s contributions.

The main novelty here, compared to our previous work~\cite{He2009} without {\it t}-channel, appears at most
backward angles, where the present model reproduces those data points much better.
\subsection{Helicity amplitudes}\label{Sec:Hel}
In Table~\ref{Tab:Amplitudes}, we present the predictions of our full model for the helicity
amplitudes and the partial decay widths of $N^*~\to~\eta N$, $\pi N$ decay channels,
for all $n$~=1 and 2 shell resonances generated by the quark model, and reported in PDG.

\begin{table*}[ht!]
\caption{{\footnotesize Helicity amplitudes and decay widths for resonances, with
$\Gamma_{\eta(\pi) N}^{PDG}=\Gamma_{tot}\cdot Br_{\eta(\pi) N}$ in PDG
~\cite{Amsler2008}. Here $\sigma$ is the
sign for $\pi N\to~\eta N$ as in Ref.~\cite{Koniuk1980}.
}}
\renewcommand\tabcolsep{0.16cm}
\begin{tabular}{ l|  c lr r|r r|rr }  \hline\hline
Resonances     & $A_{1/2}$   & $A_{1/2}^{PDG}$  &   $A_{3/2}$  &
$A_{3/2}^{PDG}$  & $\sigma\sqrt{\Gamma_{\eta N}}$ &
$(\sigma)\sqrt{\Gamma_{\eta N}^{PDG}}$ & $\sqrt{\Gamma_{\pi N}}$ &
$\sqrt{\Gamma_{\pi N}^{PDG}}$ \\ \hline
$S_{11}$(1535) & 64& 90 $\pm$ 30   &   &                   &   7.38& $ 8.87^{+1.37}_{-1.37}$        &      8.90 &     $  8.22^{+  1.59}_{-1.60}$   \\
$S_{11}$(1650) & 58& 53 $\pm$ 16   &   &                   &  -2.40&     $ 1.95^{+ 0.94}_{-1.57}$        &     12.09 &     $ 11.31^{+  1.95}_{-1.98}$  \\
$P_{11}$(1440) &-26&-65 $\pm$ 4    &   &                   &   1.62&                                     &     19.38 &     $ 13.96^{+  4.41}_{- 3.48}$   \\
$P_{11}$(1710) &-54&  9 $\pm$ 22   &   &                   &  -1.02&     $ 2.49^{+ 1.75}_{-0.88}$        &      3.19 &     $  3.87^{+  3.20}_{-1.64}$  \\
$P_{11}$(2100)&  3&               &   &                   &  -1.01&                                     &      5.77 &     $  5.34^{+  2.16}_{-2.16} $    \\
$P_{13}$(1720) &166& 18 $\pm$ 30   &-63&  -19 $\pm$ 20     &   2.44&     $ 2.83^{+ 1.04}_{-0.71}$        &     24.35 &     $  5.48^{+  2.27}_{-1.60}$  \\
$P_{13}$(1900) & 29&               &  3&                   &  -1.19&     $ 8.35^{+ 2.11}_{-2.20}$        &     13.11 &     $ 11.38^{+  2.20}_{-2.21}$  \\
$D_{13}$(1520) &-17& -24 $\pm$  9  &142&  166 $\pm\ $5     &   0.33&
$ 0.51^{+ 0.07}_{-0.08}$        &     14.26 &     $  8.31^{+0.71}_{-0.89}$   \\
$D_{13}$(1700) & -5& -18 $\pm$ 13  &  2&   -2 $\pm$ 24     &  -0.61&     $ 0.00^{+ 1.22}_{-0.00}$        &      4.90 &     $  3.16^{+  1.58}_{-1.58}$  \\
$D_{15}$(1675) & -6&  19 $\pm$  8  & -9&   15 $\pm\ $9     &  -1.88&     $ 0.00^{+ 1.28}_{-0.00}$        &      7.56 &     $  7.75^{+  0.87}_{-1.00}$   \\
$F_{15}$(1680) & 13& -15 $\pm$  6  &123&  133 $\pm$ 12     &   0.43&     $ 0.00^{+ 1.18}_{-0.00}$        &     13.61 &     $  9.37^{+  0.53}_{-0.54}$    \\
$F_{15}$(2000) & -1&               & 11&                   &  -0.36&                                     &      3.71&     $  4.00^{+  6.20}_{-2.18}$    \\
$F_{17}$(1990) &  5&   $ 30 \pm 29$            &  7&  $ 86 \pm 60$                &  -1.18&     $ 0.00^{+ 2.17}_{-0.00}$        &      7.03&     $  4.58^{+  1.55}_{-1.55}$
\\\hline \hline\end{tabular} \label{Tab:Amplitudes}
\end{table*}

Our results (Table~\ref{Tab:Amplitudes}) come out in line with those from other similar approaches
(see Tables I and II in Ref.~\cite{Capstick2000}).
%
%
%
Among the adjustable parameters given in Table~\ref{Tab: Para}, only the first five intervene
in the entries discussed here. Those parameters are close enough to the values obtained
without {\it t}-channel contribution in our previous work~\cite{He2008a}.
So, results reported in
Table~\ref{Tab:Amplitudes} are not very different from those in Table IV of Ref.~\cite{He2008a}.
Nevertheless the present results show  a general tendency to decrease the helicity amplitudes with respect to the
previous results. This feature brings the predicted values closer to those reported in PDG.
The most significant improvements concern $S_{11}$(1650) and $D_{13}$(1520), though the latter
has a rather marginal role, except in producing the right curvature of the
photoproduction polarization observables.
The partial decay widths $N^*~\to~ \pi N$ also mostly follow a decreasing behavior, leading to
significant improvements for $S_{11}$(1535) and $S_{11}$(1650). 
With respect to the five missing resonances ($P_{11}$(1899), $P_{13}$(1942), $P_{13}$(1965),
$P_{13}$(2047), $F_{15}$(1943), not shown in Table~\ref{Tab:Amplitudes}, the helicity amplitudes 
differ slightly from those in Ref.~\cite{He2008a},
and braching ratios show more sensitivity to the {\it t}-channel treatments, but variations still remain below 10\%.
Finally, none of the results display a drastic undesirable change due to the inclusion of {\it t}-channel
contributions.

\section{Summary and Conclusion}\label{Sec:Conclu}

Chiral constituent quark approach, embodying $n \leq$ 2 shell
resonances, has proven to be an appropriate formalism in interpreting
data for processes $\gamma p \to \eta p$ and $\pi^- p \to \eta n$ from
threshold up to $W \lesssim$ 2 GeV.  The effective Lagrangian
approaches, based on meson-nucleon degrees of freedom, allow also
comparable success with respect to the available data.  However, the
$\chi CQM$ offers on the one hand insights into the subnucleonic
structure of hadrons, and on the other requires (much) smaller number
of adjustable parameters, a number which increases significantly
within effective Lagrangian treatments while including an increasing
number of nucleon resonances.

Developing a formalism adequate for energies above $W \approx$ 2 GeV
implies, either the extension of the $\chi CQM$ formalism to $n >$ 2
shells, or the inclusion of {\it t}-channel contributions.  In the
present work we have adopted the second option and complemented our
$\chi CQM$ formalism with a Reggeized trajectories treatment. Such an
effort is motivated by the new data released by the
CBELSA/TAPS~\cite{Crede2009} and CLAS~\cite{Williams2009}
collaborations.  The {\it s}-channel contributions are calculated in
the quark level starting from the effective chiral Lagrangian and {\it
u}-channel is treated as degenerate.  For the {\it t}-channel a
Reggeized model is introduced at quark level.

A database, with about 4000 data points, for both $\eta$
photoproduction and $\pi^-p\to\eta n$ processes was used in the
fitting procedure, with 19 adjustable parameters.  The whole database
is well reproduced with an average of reduced $\chi^2$ of about 2.4,
except for the CLAS results~\cite{Dugger2002} published in 2002. As an
outcome, we reach the following conclusions:

\begin{itemize}
\item {The reaction mechanism in {\it s}-channel is dominated by five known resonances, namely,
$S_{11}(1535)$, $S_{11}(1650)$, $D_{13}(1520)$, $F_{15}(1680)$, and $P_{13}(1900)$, and one
new $S_{11}$ resonance with $M \approx$ 1.7 GeV. The missing resonances generated by the OGE
mechanism show no influence on the studied reactions mechanisms.}
\item {The Reggeized {\it t}-channel is indispensable in reproducing data above $W \approx$ 2 GeV.}
\item {The {\it u}-channel contributions, though treated in degenerate approximation, play significant
role in extreme forward angles for $d\sigma/d\Omega(\pi^- p\to\eta n)$.}
\item {The main coupling constant, $g_{\eta NN}$ turns out to be rather small, with a value of $\approx$ 0.3.}
\item {The extracted couplings for the $\rho$ and $\omega$ mesons come out compatible with values
known from other sources.}
\item {The mass and width of the main nucleon resonance, $S_{11}$ (1535), left as free parameters,
get values in agreement with those reported in PDG. This result emphasizes the stability of the minimization
procedure.}
\end{itemize}

To go further in such investigations, there is a real need for new
$\pi N \to \eta N$ data in the energy range covered already by
photoproduction data, corresponding to the whole range of the nucleon
resonance masses.  From theoretical side, extending the $\chi CQM$,
with {\it s}-channel embodying resonances above $n$=2 shell seems
desirable and a work in that direction is in progress. 
That approach allows including
all known resonances, even those with masses higher than 2 GeV.
Then that work is expected to provide a suitable context to test the duality principle
within our formalism.

\section*{Acknowledgements}

This project is supported by National Natural Science Foundation of
China under Grants 10905077, the Project-sponsored by SRF for ROCS,
SEM under grant HGJO90402, and the special Foundation of president
fellowship of Chinese Academy of sciences under grant YZ080425.


\begin{thebibliography}{94}
\expandafter\ifx\csname natexlab\endcsname\relax\def\natexlab#1{#1}\fi
\expandafter\ifx\csname bibnamefont\endcsname\relax
  \def\bibnamefont#1{#1}\fi
\expandafter\ifx\csname bibfnamefont\endcsname\relax
  \def\bibfnamefont#1{#1}\fi
\expandafter\ifx\csname citenamefont\endcsname\relax
  \def\citenamefont#1{#1}\fi
\expandafter\ifx\csname url\endcsname\relax
  \def\url#1{\texttt{#1}}\fi
\expandafter\ifx\csname urlprefix\endcsname\relax\def\urlprefix{URL }\fi
\providecommand{\bibinfo}[2]{#2}
\providecommand{\eprint}[2][]{\url{#2}}

\bibitem[{\citenamefont{Isgur and Karl}(1977)}]{Isgur1977}
\bibinfo{author}{\bibfnamefont{N.}~\bibnamefont{Isgur}} \bibnamefont{and}
  \bibinfo{author}{\bibfnamefont{G.}~\bibnamefont{Karl}},
  \bibinfo{journal}{Phys. Lett.} \textbf{\bibinfo{volume}{B72}},
  \bibinfo{pages}{109} (\bibinfo{year}{1977}).

\bibitem[{\citenamefont{Isgur}(2000)}]{Isgur2000a}
\bibinfo{author}{\bibfnamefont{N.}~\bibnamefont{Isgur}},
  \bibinfo{journal}{Phys. Rev.} \textbf{\bibinfo{volume}{D62}},
  \bibinfo{pages}{054026} (\bibinfo{year}{2000}).

\bibitem[{\citenamefont{Saghai and Li}(2001)}]{Saghai2001}
\bibinfo{author}{\bibfnamefont{B.}~\bibnamefont{Saghai}} \bibnamefont{and}
  \bibinfo{author}{\bibfnamefont{Z.}~\bibnamefont{Li}}, \bibinfo{journal}{Eur.
  Phy. J.} \textbf{\bibinfo{volume}{A11}}, \bibinfo{pages}{217}
  (\bibinfo{year}{2001}).

\bibitem[{\citenamefont{Chizma and Karl}(2003)}]{Chizma2003}
\bibinfo{author}{\bibfnamefont{J.}~\bibnamefont{Chizma}} \bibnamefont{and}
  \bibinfo{author}{\bibfnamefont{G.}~\bibnamefont{Karl}},
  \bibinfo{journal}{Phys. Rev.} \textbf{\bibinfo{volume}{D68}},
  \bibinfo{pages}{054007} (\bibinfo{year}{2003}).

\bibitem[{\citenamefont{He and Dong}(2003{\natexlab{a}})}]{He2003a}
\bibinfo{author}{\bibfnamefont{J.}~\bibnamefont{He}} \bibnamefont{and}
  \bibinfo{author}{\bibfnamefont{Y.-B.} \bibnamefont{Dong}},
  \bibinfo{journal}{Nucl. Phys.} \textbf{\bibinfo{volume}{A725}},
  \bibinfo{pages}{201} (\bibinfo{year}{2003}{\natexlab{a}}).

\bibitem[{\citenamefont{He and Dong}(2003{\natexlab{b}})}]{He2003b}
\bibinfo{author}{\bibfnamefont{J.}~\bibnamefont{He}} \bibnamefont{and}
  \bibinfo{author}{\bibfnamefont{Y.-B.} \bibnamefont{Dong}},
  \bibinfo{journal}{J. Phys.} \textbf{\bibinfo{volume}{G29}},
  \bibinfo{pages}{2737} (\bibinfo{year}{2003}{\natexlab{b}}).

\bibitem[{\citenamefont{Capstick and Roberts}(2004)}]{Capstick2004a}
\bibinfo{author}{\bibfnamefont{S.}~\bibnamefont{Capstick}} \bibnamefont{and}
  \bibinfo{author}{\bibfnamefont{W.}~\bibnamefont{Roberts}},
  \bibinfo{journal}{Fizika} \textbf{\bibinfo{volume}{B13}},
  \bibinfo{pages}{271} (\bibinfo{year}{2004}).

\bibitem[{\citenamefont{Saghai and Li}(2010)}]{Saghai2010}
\bibinfo{author}{\bibfnamefont{B.}~\bibnamefont{Saghai}} \bibnamefont{and}
  \bibinfo{author}{\bibfnamefont{Z.}~\bibnamefont{Li}}, \bibinfo{journal}{Few
  Body Syst.} \textbf{\bibinfo{volume}{47}}, \bibinfo{pages}{105}
  (\bibinfo{year}{2010}).

\bibitem[{\citenamefont{An and Zou}(2009{\natexlab{a}})}]{An2009}
\bibinfo{author}{\bibfnamefont{C.~S.} \bibnamefont{An}} \bibnamefont{and}
  \bibinfo{author}{\bibfnamefont{B.~S.} \bibnamefont{Zou}},
  \bibinfo{journal}{Eur. Phys. J.} \textbf{\bibinfo{volume}{A39}},
  \bibinfo{pages}{195} (\bibinfo{year}{2009}{\natexlab{a}}).

\bibitem[{\citenamefont{An and Zou}(2009{\natexlab{b}})}]{An2009a}
\bibinfo{author}{\bibfnamefont{C.-S.} \bibnamefont{An}} \bibnamefont{and}
  \bibinfo{author}{\bibfnamefont{B.-S.} \bibnamefont{Zou}},
  \bibinfo{journal}{Sci. Sin.} \textbf{\bibinfo{volume}{G52}},
  \bibinfo{pages}{1452} (\bibinfo{year}{2009}{\natexlab{b}}).

\bibitem[{\citenamefont{Deinet et~al.}(1969)}]{Deinet1969}
\bibinfo{author}{\bibfnamefont{W.}~\bibnamefont{Deinet}} \bibnamefont{et~al.},
  \bibinfo{journal}{Nucl. Phys.} \textbf{\bibinfo{volume}{B11}},
  \bibinfo{pages}{495} (\bibinfo{year}{1969}).

\bibitem[{\citenamefont{Richards et~al.}(1970)}]{Richards1970}
\bibinfo{author}{\bibfnamefont{W.~B.} \bibnamefont{Richards}}
  \bibnamefont{et~al.}, \bibinfo{journal}{Phys. Rev. D}
  \textbf{\bibinfo{volume}{1}}, \bibinfo{pages}{10} (\bibinfo{year}{1970}).

\bibitem[{\citenamefont{Debenham et~al.}(1975)}]{Debenham1975}
\bibinfo{author}{\bibfnamefont{N.~C.} \bibnamefont{Debenham}}
  \bibnamefont{et~al.}, \bibinfo{journal}{Phys. Rev.}
  \textbf{\bibinfo{volume}{D12}}, \bibinfo{pages}{2545} (\bibinfo{year}{1975}).

\bibitem[{\citenamefont{Brown et~al.}(1979)}]{Brown1979}
\bibinfo{author}{\bibfnamefont{R.~M.} \bibnamefont{Brown}}
  \bibnamefont{et~al.}, \bibinfo{journal}{Nucl. Phys.}
  \textbf{\bibinfo{volume}{B153}}, \bibinfo{pages}{89} (\bibinfo{year}{1979}).

\bibitem[{\citenamefont{Clajus and Nefkens}(1992)}]{Clajus1992}
\bibinfo{author}{\bibfnamefont{M.}~\bibnamefont{Clajus}} \bibnamefont{and}
  \bibinfo{author}{\bibfnamefont{B.~M.~K.} \bibnamefont{Nefkens}},
  \bibinfo{journal}{PiN Newslett.} \textbf{\bibinfo{volume}{7}},
  \bibinfo{pages}{76} (\bibinfo{year}{1992}).

\bibitem[{\citenamefont{Morrison}(1999)}]{Morrison1999}
\bibinfo{author}{\bibfnamefont{T.~W.} \bibnamefont{Morrison}}, Ph.D. thesis,
  \bibinfo{school}{The George Washington University} (\bibinfo{year}{1999}).

\bibitem[{\citenamefont{Prakhov et~al.}(2005)}]{Prakhov2005}
\bibinfo{author}{\bibfnamefont{S.}~\bibnamefont{Prakhov}} \bibnamefont{et~al.},
  \bibinfo{journal}{Phys. Rev.} \textbf{\bibinfo{volume}{C72}},
  \bibinfo{pages}{015203} (\bibinfo{year}{2005}).

\bibitem[{\citenamefont{Dugger et~al.}(2002)}]{Dugger2002}
\bibinfo{author}{\bibfnamefont{M.~B.} \bibnamefont{Dugger}}
  \bibnamefont{et~al.} (\bibinfo{collaboration}{CLAS}), \bibinfo{journal}{Phys.
  Rev. Lett.} \textbf{\bibinfo{volume}{89}}, \bibinfo{pages}{222002}
  (\bibinfo{year}{2002}).

\bibitem[{\citenamefont{Crede et~al.}(2005)}]{Crede2005}
\bibinfo{author}{\bibfnamefont{V.}~\bibnamefont{Crede}} \bibnamefont{et~al.}
  (\bibinfo{collaboration}{CB-ELSA Collaboration}), \bibinfo{journal}{Phys.
  Rev. Lett.} \textbf{\bibinfo{volume}{94}}, \bibinfo{eid}{012004}
  (pages~\bibinfo{numpages}{5}) (\bibinfo{year}{2005}).

\bibitem[{\citenamefont{Nakabayashi et~al.}(2006)}]{Nakabayashi2006}
\bibinfo{author}{\bibfnamefont{T.}~\bibnamefont{Nakabayashi}}
  \bibnamefont{et~al.}, \bibinfo{journal}{Phys. Rev. }
  \textbf{\bibinfo{volume}{C74}}, \bibinfo{eid}{035202}
  (pages~\bibinfo{numpages}{7}) (\bibinfo{year}{2006}).

\bibitem[{\citenamefont{Bartalini et~al.}(2007)}]{Bartalini2007}
\bibinfo{author}{\bibfnamefont{O.}~\bibnamefont{Bartalini}}
  \bibnamefont{et~al.} (\bibinfo{collaboration}{The GRAAL}),
  \bibinfo{journal}{Eur. Phys. J.} \textbf{\bibinfo{volume}{A33}},
  \bibinfo{pages}{169} (\bibinfo{year}{2007}).

\bibitem[{\citenamefont{Elsner et~al.}(2007)}]{Elsner2007}
\bibinfo{author}{\bibfnamefont{D.}~\bibnamefont{Elsner}} \bibnamefont{et~al.}
  (\bibinfo{collaboration}{CBELSA}), \bibinfo{journal}{Eur. Phys. J.}
  \textbf{\bibinfo{volume}{A33}}, \bibinfo{pages}{147} (\bibinfo{year}{2007}).

\bibitem[{\citenamefont{Crede et~al.}(2009)}]{Crede2009}
\bibinfo{author}{\bibfnamefont{V.}~\bibnamefont{Crede}} \bibnamefont{et~al.}
  (\bibinfo{collaboration}{CBELSA/TAPS}), \bibinfo{journal}{Phys. Rev.}
  \textbf{\bibinfo{volume}{C80}}, \bibinfo{pages}{055202}
  (\bibinfo{year}{2009}).

\bibitem[{\citenamefont{Williams et~al.}(2009)}]{Williams2009}
\bibinfo{author}{\bibfnamefont{M.}~\bibnamefont{Williams}} \bibnamefont{et~al.}
  (\bibinfo{collaboration}{CLAS}), \bibinfo{journal}{Phys. Rev.}
  \textbf{\bibinfo{volume}{C80}}, \bibinfo{pages}{045213}
  (\bibinfo{year}{2009}).

\bibitem[{\citenamefont{Feuster and Mosel}(1998)}]{Feuster1998}
\bibinfo{author}{\bibfnamefont{T.}~\bibnamefont{Feuster}} \bibnamefont{and}
  \bibinfo{author}{\bibfnamefont{U.}~\bibnamefont{Mosel}},
  \bibinfo{journal}{Phys. Rev.} \textbf{\bibinfo{volume}{C58}},
  \bibinfo{pages}{457} (\bibinfo{year}{1998}).

\bibitem[{\citenamefont{Feuster and Mosel}(1999)}]{Feuster1999}
\bibinfo{author}{\bibfnamefont{T.}~\bibnamefont{Feuster}} \bibnamefont{and}
  \bibinfo{author}{\bibfnamefont{U.}~\bibnamefont{Mosel}},
  \bibinfo{journal}{Phys. Rev.} \textbf{\bibinfo{volume}{C59}},
  \bibinfo{pages}{460} (\bibinfo{year}{1999}).

\bibitem[{\citenamefont{Penner}(2002)}]{Penner2002}
\bibinfo{author}{\bibfnamefont{G.}~\bibnamefont{Penner}}, Ph.D. thesis,
  \bibinfo{school}{Universit\"{a}t Giessen} (\bibinfo{year}{2002}).

\bibitem[{\citenamefont{Penner and Mosel}(2002)}]{Penner2002d}
\bibinfo{author}{\bibfnamefont{G.}~\bibnamefont{Penner}} \bibnamefont{and}
  \bibinfo{author}{\bibfnamefont{U.}~\bibnamefont{Mosel}},
  \bibinfo{journal}{Phys. Rev.} \textbf{\bibinfo{volume}{C66}},
  \bibinfo{pages}{055211} (\bibinfo{year}{2002}).

\bibitem[{\citenamefont{Anisovich et~al.}(2005)}]{Anisovich2005}
\bibinfo{author}{\bibfnamefont{A.~V.} \bibnamefont{Anisovich}}
  \bibnamefont{et~al.}, \bibinfo{journal}{Eur. Phys. J.}
  \textbf{\bibinfo{volume}{A25}}, \bibinfo{pages}{427} (\bibinfo{year}{2005}).

\bibitem[{\citenamefont{Sarantsev et~al.}(2005)\citenamefont{Sarantsev,
  Nikonov, Anisovich, Klempt, and Thoma}}]{Sarantsev2005}
\bibinfo{author}{\bibfnamefont{A.~V.} \bibnamefont{Sarantsev}},
  \bibinfo{author}{\bibfnamefont{V.~A.} \bibnamefont{Nikonov}},
  \bibinfo{author}{\bibfnamefont{A.~V.} \bibnamefont{Anisovich}},
  \bibinfo{author}{\bibfnamefont{E.}~\bibnamefont{Klempt}}, \bibnamefont{and}
  \bibinfo{author}{\bibfnamefont{U.}~\bibnamefont{Thoma}},
  \bibinfo{journal}{Eur. Phys. J.} \textbf{\bibinfo{volume}{A25}},
  \bibinfo{pages}{441} (\bibinfo{year}{2005}).

\bibitem[{\citenamefont{Batinic et~al.}(1997)}]{Batinic1997}
\bibinfo{author}{\bibfnamefont{M.}~\bibnamefont{Batinic}} \bibnamefont{et~al.}
  (\bibinfo{year}{1997}), \eprint{nucl-th/9703023}.

\bibitem[{\citenamefont{Ceci et~al.}(2006)\citenamefont{Ceci, Svarc, and
  Zauner}}]{Ceci2006}
\bibinfo{author}{\bibfnamefont{S.}~\bibnamefont{Ceci}},
  \bibinfo{author}{\bibfnamefont{A.}~\bibnamefont{Svarc}}, \bibnamefont{and}
  \bibinfo{author}{\bibfnamefont{B.}~\bibnamefont{Zauner}},
  \bibinfo{journal}{Phys. Rev. Lett.} \textbf{\bibinfo{volume}{97}},
  \bibinfo{pages}{062002} (\bibinfo{year}{2006}).

\bibitem[{\citenamefont{Arndt et~al.}(2007)\citenamefont{Arndt, Briscoe,
  Strakovsky, and Workman}}]{Arndt2007}
\bibinfo{author}{\bibfnamefont{R.~A.} \bibnamefont{Arndt}},
  \bibinfo{author}{\bibfnamefont{W.~J.} \bibnamefont{Briscoe}},
  \bibinfo{author}{\bibfnamefont{I.~I.} \bibnamefont{Strakovsky}},
  \bibnamefont{and} \bibinfo{author}{\bibfnamefont{R.~L.}
  \bibnamefont{Workman}}, \bibinfo{journal}{Int. J. Mod. Phys.}
  \textbf{\bibinfo{volume}{A22}}, \bibinfo{pages}{349} (\bibinfo{year}{2007}).

\bibitem[{\citenamefont{Vrana et~al.}(2000)\citenamefont{Vrana, Dytman, and
  Lee}}]{Vrana2000}
\bibinfo{author}{\bibfnamefont{T.~P.} \bibnamefont{Vrana}},
  \bibinfo{author}{\bibfnamefont{S.~A.} \bibnamefont{Dytman}},
  \bibnamefont{and} \bibinfo{author}{\bibfnamefont{T.~S.~H.}
  \bibnamefont{Lee}}, \bibinfo{journal}{Phys. Rept.}
  \textbf{\bibinfo{volume}{328}}, \bibinfo{pages}{181} (\bibinfo{year}{2000}).

\bibitem[{\citenamefont{Matsuyama et~al.}(2007)\citenamefont{Matsuyama, Sato,
  and Lee}}]{Matsuyama2007}
\bibinfo{author}{\bibfnamefont{A.}~\bibnamefont{Matsuyama}},
  \bibinfo{author}{\bibfnamefont{T.}~\bibnamefont{Sato}}, \bibnamefont{and}
  \bibinfo{author}{\bibfnamefont{T.~S.~H.} \bibnamefont{Lee}},
  \bibinfo{journal}{Phys. Rept.} \textbf{\bibinfo{volume}{439}},
  \bibinfo{pages}{193} (\bibinfo{year}{2007}).

\bibitem[{\citenamefont{Chen et~al.}(2003)\citenamefont{Chen, Kamalov, Yang,
  Drechsel, and Tiator}}]{Chen2003}
\bibinfo{author}{\bibfnamefont{G.-Y.} \bibnamefont{Chen}},
  \bibinfo{author}{\bibfnamefont{S.}~\bibnamefont{Kamalov}},
  \bibinfo{author}{\bibfnamefont{S.~N.} \bibnamefont{Yang}},
  \bibinfo{author}{\bibfnamefont{D.}~\bibnamefont{Drechsel}}, \bibnamefont{and}
  \bibinfo{author}{\bibfnamefont{L.}~\bibnamefont{Tiator}},
  \bibinfo{journal}{Nucl. Phys.} \textbf{\bibinfo{volume}{A723}},
  \bibinfo{pages}{447} (\bibinfo{year}{2003}).

\bibitem[{\citenamefont{Chen et~al.}(2007)\citenamefont{Chen, Dong, Giannini,
  and Santopinto}}]{Chen2007a}
\bibinfo{author}{\bibfnamefont{D.~Y.} \bibnamefont{Chen}},
  \bibinfo{author}{\bibfnamefont{Y.~B.} \bibnamefont{Dong}},
  \bibinfo{author}{\bibfnamefont{M.~M.} \bibnamefont{Giannini}},
  \bibnamefont{and}
  \bibinfo{author}{\bibfnamefont{E.}~\bibnamefont{Santopinto}},
  \bibinfo{journal}{Nucl. Phys.} \textbf{\bibinfo{volume}{A782}},
  \bibinfo{pages}{62} (\bibinfo{year}{2007}).

\bibitem[{\citenamefont{Gasparyan et~al.}(2003)\citenamefont{Gasparyan,
  Haidenbauer, Hanhart, and Speth}}]{Gasparyan2003}
\bibinfo{author}{\bibfnamefont{A.}~\bibnamefont{Gasparyan}},
  \bibinfo{author}{\bibfnamefont{J.}~\bibnamefont{Haidenbauer}},
  \bibinfo{author}{\bibfnamefont{C.}~\bibnamefont{Hanhart}}, \bibnamefont{and}
  \bibinfo{author}{\bibfnamefont{J.}~\bibnamefont{Speth}},
  \bibinfo{journal}{Phys. Rev} \textbf{\bibinfo{volume}{C68}},
  \bibinfo{pages}{045207} (\bibinfo{year}{2003}).

\bibitem[{\citenamefont{Julia-Diaz et~al.}(2006)\citenamefont{Julia-Diaz,
  Saghai, Lee, and Tabakin}}]{Julia-Diaz2006}
\bibinfo{author}{\bibfnamefont{B.}~\bibnamefont{Julia-Diaz}},
  \bibinfo{author}{\bibfnamefont{B.}~\bibnamefont{Saghai}},
  \bibinfo{author}{\bibfnamefont{T.-S.} \bibnamefont{Lee}}, \bibnamefont{and}
  \bibinfo{author}{\bibfnamefont{F.}~\bibnamefont{Tabakin}},
  \bibinfo{journal}{Phys.Rev.} \textbf{\bibinfo{volume}{C73}},
  \bibinfo{pages}{055204} (\bibinfo{year}{2006}).

\bibitem[{\citenamefont{Julia-Diaz et~al.}(2007)\citenamefont{Julia-Diaz, Lee,
  Matsuyama, and Sato}}]{Julia-Diaz2007}
\bibinfo{author}{\bibfnamefont{B.}~\bibnamefont{Julia-Diaz}},
  \bibinfo{author}{\bibfnamefont{T.~S.~H.} \bibnamefont{Lee}},
  \bibinfo{author}{\bibfnamefont{A.}~\bibnamefont{Matsuyama}},
  \bibnamefont{and} \bibinfo{author}{\bibfnamefont{T.}~\bibnamefont{Sato}},
  \bibinfo{journal}{Phys. Rev.} \textbf{\bibinfo{volume}{C76}},
  \bibinfo{pages}{065201} (\bibinfo{year}{2007}).

\bibitem[{\citenamefont{Durand et~al.}(2008)\citenamefont{Durand, Julia-Diaz,
  Lee, Saghai, and Sato}}]{Durand2008}
\bibinfo{author}{\bibfnamefont{J.}~\bibnamefont{Durand}},
  \bibinfo{author}{\bibfnamefont{B.}~\bibnamefont{Julia-Diaz}},
  \bibinfo{author}{\bibfnamefont{T.~S.~H.} \bibnamefont{Lee}},
  \bibinfo{author}{\bibfnamefont{B.}~\bibnamefont{Saghai}}, \bibnamefont{and}
  \bibinfo{author}{\bibfnamefont{T.}~\bibnamefont{Sato}},
  \bibinfo{journal}{Phys. Rev.} \textbf{\bibinfo{volume}{C78}},
  \bibinfo{pages}{025204} (\bibinfo{year}{2008}).

\bibitem[{\citenamefont{Durand et~al.}(2009)\citenamefont{Durand, Julia-Diaz,
  Lee, Saghai, and Sato}}]{Durand2009}
\bibinfo{author}{\bibfnamefont{J.}~\bibnamefont{Durand}},
  \bibinfo{author}{\bibfnamefont{B.}~\bibnamefont{Julia-Diaz}},
  \bibinfo{author}{\bibfnamefont{T.~S.~H.} \bibnamefont{Lee}},
  \bibinfo{author}{\bibfnamefont{B.}~\bibnamefont{Saghai}}, \bibnamefont{and}
  \bibinfo{author}{\bibfnamefont{T.}~\bibnamefont{Sato}},
  \bibinfo{journal}{Int. J. Mod. Phys.} \textbf{\bibinfo{volume}{A24}},
  \bibinfo{pages}{553} (\bibinfo{year}{2009}).

\bibitem[{\citenamefont{Chiang et~al.}(2003)\citenamefont{Chiang, Yang, Tiator,
  Vanderhaeghen, and Drechsel}}]{Chiang2003}
\bibinfo{author}{\bibfnamefont{W.-T.} \bibnamefont{Chiang}},
  \bibinfo{author}{\bibfnamefont{S.~N.} \bibnamefont{Yang}},
  \bibinfo{author}{\bibfnamefont{L.}~\bibnamefont{Tiator}},
  \bibinfo{author}{\bibfnamefont{M.}~\bibnamefont{Vanderhaeghen}},
  \bibnamefont{and} \bibinfo{author}{\bibfnamefont{D.}~\bibnamefont{Drechsel}},
  \bibinfo{journal}{Phys. Rev.} \textbf{\bibinfo{volume}{C68}},
  \bibinfo{pages}{045202} (\bibinfo{year}{2003}).

\bibitem[{\citenamefont{Chiang et~al.}(2004)\citenamefont{Chiang, Saghai,
  Tabakin, and Lee}}]{Chiang2004}
\bibinfo{author}{\bibfnamefont{W.-T.} \bibnamefont{Chiang}},
  \bibinfo{author}{\bibfnamefont{B.}~\bibnamefont{Saghai}},
  \bibinfo{author}{\bibfnamefont{F.}~\bibnamefont{Tabakin}}, \bibnamefont{and}
  \bibinfo{author}{\bibfnamefont{T.~S.~H.} \bibnamefont{Lee}},
  \bibinfo{journal}{Phys. Rev.} \textbf{\bibinfo{volume}{C69}},
  \bibinfo{pages}{065208} (\bibinfo{year}{2004}).

\bibitem[{\citenamefont{Zhao et~al.}(2002)\citenamefont{Zhao, Al-Khalili, Li,
  and Workman}}]{Zhao2002}
\bibinfo{author}{\bibfnamefont{Q.}~\bibnamefont{Zhao}},
  \bibinfo{author}{\bibfnamefont{J.~S.} \bibnamefont{Al-Khalili}},
  \bibinfo{author}{\bibfnamefont{Z.~P.} \bibnamefont{Li}}, \bibnamefont{and}
  \bibinfo{author}{\bibfnamefont{R.~L.} \bibnamefont{Workman}},
  \bibinfo{journal}{Phys. Rev.} \textbf{\bibinfo{volume}{C65}},
  \bibinfo{pages}{065204} (\bibinfo{year}{2002}).

\bibitem[{\citenamefont{He et~al.}(2008{\natexlab{a}})\citenamefont{He, Saghai,
  Li, Zhao, and Durand}}]{He2008}
\bibinfo{author}{\bibfnamefont{J.}~\bibnamefont{He}},
  \bibinfo{author}{\bibfnamefont{B.}~\bibnamefont{Saghai}},
  \bibinfo{author}{\bibfnamefont{Z.}~\bibnamefont{Li}},
  \bibinfo{author}{\bibfnamefont{Q.}~\bibnamefont{Zhao}}, \bibnamefont{and}
  \bibinfo{author}{\bibfnamefont{J.}~\bibnamefont{Durand}},
  \bibinfo{journal}{Eur. Phys. J.} \textbf{\bibinfo{volume}{A35}},
  \bibinfo{pages}{321} (\bibinfo{year}{2008}{\natexlab{a}}).

\bibitem[{\citenamefont{He et~al.}(2008{\natexlab{b}})\citenamefont{He, Saghai,
  and Li}}]{He2008a}
\bibinfo{author}{\bibfnamefont{J.}~\bibnamefont{He}},
  \bibinfo{author}{\bibfnamefont{B.}~\bibnamefont{Saghai}}, \bibnamefont{and}
  \bibinfo{author}{\bibfnamefont{Z.}~\bibnamefont{Li}}, \bibinfo{journal}{Phys.
  Rev.} \textbf{\bibinfo{volume}{C78}}, \bibinfo{pages}{035204}
  (\bibinfo{year}{2008}{\natexlab{b}}).

\bibitem[{\citenamefont{He and Saghai}(2009)}]{He2009}
\bibinfo{author}{\bibfnamefont{J.}~\bibnamefont{He}} \bibnamefont{and}
  \bibinfo{author}{\bibfnamefont{B.}~\bibnamefont{Saghai}},
  \bibinfo{journal}{Phys. Rev.} \textbf{\bibinfo{volume}{C80}},
  \bibinfo{pages}{015207} (\bibinfo{year}{2009}).

\bibitem[{\citenamefont{Zhong et~al.}(2007)\citenamefont{Zhong, Zhao, He, and
  Saghai}}]{Zhong2007}
\bibinfo{author}{\bibfnamefont{X.-H.} \bibnamefont{Zhong}},
  \bibinfo{author}{\bibfnamefont{Q.}~\bibnamefont{Zhao}},
  \bibinfo{author}{\bibfnamefont{J.}~\bibnamefont{He}}, \bibnamefont{and}
  \bibinfo{author}{\bibfnamefont{B.}~\bibnamefont{Saghai}},
  \bibinfo{journal}{Phys. Rev.} \textbf{\bibinfo{volume}{C76}},
  \bibinfo{pages}{065205} (\bibinfo{year}{2007}).

\bibitem[{\citenamefont{Collins}(1997)}]{Collins}
\bibinfo{author}{\bibfnamefont{P.~D.~B.} \bibnamefont{Collins}},
  \emph{\bibinfo{title}{{An Introduction to Regge Theory and High-Energy
  Physics}}} (\bibinfo{year}{1997}), \bibinfo{note}{{Cambridge University
  Press}}.

\bibitem[{\citenamefont{Saghai and Tabakin}(1996)}]{Saghai1996}
\bibinfo{author}{\bibfnamefont{B.}~\bibnamefont{Saghai}} \bibnamefont{and}
  \bibinfo{author}{\bibfnamefont{F.}~\bibnamefont{Tabakin}},
  \bibinfo{journal}{Phys. Rev.} \textbf{\bibinfo{volume}{C53}},
  \bibinfo{pages}{66} (\bibinfo{year}{1996}).

\bibitem[{\citenamefont{Guidal et~al.}(1997)\citenamefont{Guidal, Laget, and
  Vanderhaeghen}}]{Guidal1997a}
\bibinfo{author}{\bibfnamefont{M.}~\bibnamefont{Guidal}},
  \bibinfo{author}{\bibfnamefont{J.~M.} \bibnamefont{Laget}}, \bibnamefont{and}
  \bibinfo{author}{\bibfnamefont{M.}~\bibnamefont{Vanderhaeghen}},
  \bibinfo{journal}{Nucl. Phys.} \textbf{\bibinfo{volume}{A627}},
  \bibinfo{pages}{645} (\bibinfo{year}{1997}).

\bibitem[{\citenamefont{Sibirtsev et~al.}(2007)}]{Sibirtsev2007a}
\bibinfo{author}{\bibfnamefont{A.}~\bibnamefont{Sibirtsev}}
  \bibnamefont{et~al.}, \bibinfo{journal}{Eur. Phys. J.}
  \textbf{\bibinfo{volume}{A34}}, \bibinfo{pages}{49} (\bibinfo{year}{2007}).

\bibitem[{\citenamefont{Sibirtsev
  et~al.}(2009{\natexlab{a}})\citenamefont{Sibirtsev, Haidenbauer, Krewald,
  Meissner, and Thomas}}]{Sibirtsev2009}
\bibinfo{author}{\bibfnamefont{A.}~\bibnamefont{Sibirtsev}},
  \bibinfo{author}{\bibfnamefont{J.}~\bibnamefont{Haidenbauer}},
  \bibinfo{author}{\bibfnamefont{S.}~\bibnamefont{Krewald}},
  \bibinfo{author}{\bibfnamefont{U.~G.} \bibnamefont{Meissner}},
  \bibnamefont{and} \bibinfo{author}{\bibfnamefont{A.~W.}
  \bibnamefont{Thomas}}, \bibinfo{journal}{Eur. Phys. J.}
  \textbf{\bibinfo{volume}{A41}}, \bibinfo{pages}{71}
  (\bibinfo{year}{2009}{\natexlab{a}}).

\bibitem[{\citenamefont{Sibirtsev
  et~al.}(2009{\natexlab{b}})\citenamefont{Sibirtsev, Haidenbauer, Huang,
  Krewald, and Meissner}}]{Sibirtsev2009a}
\bibinfo{author}{\bibfnamefont{A.}~\bibnamefont{Sibirtsev}},
  \bibinfo{author}{\bibfnamefont{J.}~\bibnamefont{Haidenbauer}},
  \bibinfo{author}{\bibfnamefont{F.}~\bibnamefont{Huang}},
  \bibinfo{author}{\bibfnamefont{S.}~\bibnamefont{Krewald}}, \bibnamefont{and}
  \bibinfo{author}{\bibfnamefont{U.~G.} \bibnamefont{Meissner}},
  \bibinfo{journal}{Eur. Phys. J.} \textbf{\bibinfo{volume}{A40}},
  \bibinfo{pages}{65} (\bibinfo{year}{2009}{\natexlab{b}}).

\bibitem[{\citenamefont{Guidal et~al.}(2003)\citenamefont{Guidal, Laget, and
  Vanderhaeghen}}]{Guidal2003}
\bibinfo{author}{\bibfnamefont{M.}~\bibnamefont{Guidal}},
  \bibinfo{author}{\bibfnamefont{J.~M.} \bibnamefont{Laget}}, \bibnamefont{and}
  \bibinfo{author}{\bibfnamefont{M.}~\bibnamefont{Vanderhaeghen}},
  \bibinfo{journal}{Phys. Rev.} \textbf{\bibinfo{volume}{C68}},
  \bibinfo{pages}{058201} (\bibinfo{year}{2003}).

\bibitem[{\citenamefont{Corthals et~al.}(2006)\citenamefont{Corthals,
  Ryckebusch, and Van~Cauteren}}]{Corthals2006}
\bibinfo{author}{\bibfnamefont{T.}~\bibnamefont{Corthals}},
  \bibinfo{author}{\bibfnamefont{J.}~\bibnamefont{Ryckebusch}},
  \bibnamefont{and}
  \bibinfo{author}{\bibfnamefont{T.}~\bibnamefont{Van~Cauteren}},
  \bibinfo{journal}{Phys. Rev.} \textbf{\bibinfo{volume}{C73}},
  \bibinfo{pages}{045207} (\bibinfo{year}{2006}).

\bibitem[{\citenamefont{Corthals et~al.}(2007)\citenamefont{Corthals, Ireland,
  Van~Cauteren, and Ryckebusch}}]{Corthals2007}
\bibinfo{author}{\bibfnamefont{T.}~\bibnamefont{Corthals}},
  \bibinfo{author}{\bibfnamefont{D.~G.} \bibnamefont{Ireland}},
  \bibinfo{author}{\bibfnamefont{T.}~\bibnamefont{Van~Cauteren}},
  \bibnamefont{and}
  \bibinfo{author}{\bibfnamefont{J.}~\bibnamefont{Ryckebusch}},
  \bibinfo{journal}{Phys. Rev.} \textbf{\bibinfo{volume}{C75}},
  \bibinfo{pages}{045204} (\bibinfo{year}{2007}).

\bibitem[{\citenamefont{Vancraeyveld et~al.}(2009)\citenamefont{Vancraeyveld,
  De~Cruz, Ryckebusch, and Van~Cauteren}}]{Vancraeyveld2009a}
\bibinfo{author}{\bibfnamefont{P.}~\bibnamefont{Vancraeyveld}},
  \bibinfo{author}{\bibfnamefont{L.}~\bibnamefont{De~Cruz}},
  \bibinfo{author}{\bibfnamefont{J.}~\bibnamefont{Ryckebusch}},
  \bibnamefont{and}
  \bibinfo{author}{\bibfnamefont{T.}~\bibnamefont{Van~Cauteren}},
  \bibinfo{journal}{Phys. Lett.} \textbf{\bibinfo{volume}{B681}},
  \bibinfo{pages}{428} (\bibinfo{year}{2009}).

\bibitem[{\citenamefont{Vancraeyveld et~al.}(2010)\citenamefont{Vancraeyveld,
  De~Cruz, Ryckebusch, and Van~Cauteren}}]{Vancraeyveld2010}
\bibinfo{author}{\bibfnamefont{P.}~\bibnamefont{Vancraeyveld}},
  \bibinfo{author}{\bibfnamefont{L.}~\bibnamefont{De~Cruz}},
  \bibinfo{author}{\bibfnamefont{J.}~\bibnamefont{Ryckebusch}},
  \bibnamefont{and}
  \bibinfo{author}{\bibfnamefont{T.}~\bibnamefont{Van~Cauteren}},
  \bibinfo{journal}{EPJ Web Conf.} \textbf{\bibinfo{volume}{3}},
  \bibinfo{pages}{03013} (\bibinfo{year}{2010}).

\bibitem[{\citenamefont{Manohar and Georgi}(1984)}]{Manohar1984}
\bibinfo{author}{\bibfnamefont{A.}~\bibnamefont{Manohar}} \bibnamefont{and}
  \bibinfo{author}{\bibfnamefont{H.}~\bibnamefont{Georgi}},
  \bibinfo{journal}{Nucl. Phys.} \textbf{\bibinfo{volume}{B234}},
  \bibinfo{pages}{189} (\bibinfo{year}{1984}).

\bibitem[{\citenamefont{Li et~al.}(1997)\citenamefont{Li, Ye, and Lu}}]{Li1997}
\bibinfo{author}{\bibfnamefont{Z.-p.} \bibnamefont{Li}},
  \bibinfo{author}{\bibfnamefont{H.-x.} \bibnamefont{Ye}}, \bibnamefont{and}
  \bibinfo{author}{\bibfnamefont{M.-h.} \bibnamefont{Lu}},
  \bibinfo{journal}{Phys. Rev.} \textbf{\bibinfo{volume}{C56}},
  \bibinfo{pages}{1099} (\bibinfo{year}{1997}).

\bibitem[{\citenamefont{Chew et~al.}(1957)\citenamefont{Chew, Goldberger, Low,
  and Nambu}}]{Chew1957}
\bibinfo{author}{\bibfnamefont{G.~F.} \bibnamefont{Chew}},
  \bibinfo{author}{\bibfnamefont{M.~L.} \bibnamefont{Goldberger}},
  \bibinfo{author}{\bibfnamefont{F.~E.} \bibnamefont{Low}}, \bibnamefont{and}
  \bibinfo{author}{\bibfnamefont{Y.}~\bibnamefont{Nambu}},
  \bibinfo{journal}{Phys. Rev.} \textbf{\bibinfo{volume}{106}},
  \bibinfo{pages}{1345} (\bibinfo{year}{1957}).

\bibitem[{\citenamefont{Copley et~al.}(1969)\citenamefont{Copley, Karl, and
  Obryk}}]{Copley1969}
\bibinfo{author}{\bibfnamefont{L.~A.} \bibnamefont{Copley}},
  \bibinfo{author}{\bibfnamefont{G.}~\bibnamefont{Karl}}, \bibnamefont{and}
  \bibinfo{author}{\bibfnamefont{E.}~\bibnamefont{Obryk}},
  \bibinfo{journal}{Nucl. Phys.} \textbf{\bibinfo{volume}{B13}},
  \bibinfo{pages}{303} (\bibinfo{year}{1969}).

\bibitem[{\citenamefont{Licht and Pagnamenta}(1970)}]{Licht1970}
\bibinfo{author}{\bibfnamefont{A.~L.} \bibnamefont{Licht}} \bibnamefont{and}
  \bibinfo{author}{\bibfnamefont{A.}~\bibnamefont{Pagnamenta}},
  \bibinfo{journal}{Phys. Rev.} \textbf{\bibinfo{volume}{D2}},
  \bibinfo{pages}{1150} (\bibinfo{year}{1970}).

\bibitem[{\citenamefont{Amsler et~al.}(2008)}]{Amsler2008}
\bibinfo{author}{\bibfnamefont{C.}~\bibnamefont{Amsler}} \bibnamefont{et~al.}
  (\bibinfo{collaboration}{Particle Data Group}), \bibinfo{journal}{Phys.
  Lett.} \textbf{\bibinfo{volume}{B667}}, \bibinfo{pages}{1}
  (\bibinfo{year}{2008}).

\bibitem[{\citenamefont{Krusche et~al.}(1995)}]{Krusche1995}
\bibinfo{author}{\bibfnamefont{B.}~\bibnamefont{Krusche}} \bibnamefont{et~al.},
  \bibinfo{journal}{Phys. Rev. Lett.} \textbf{\bibinfo{volume}{74}},
  \bibinfo{pages}{3736} (\bibinfo{year}{1995}).

\bibitem[{\citenamefont{Bock et~al.}(1998)}]{Bock1998}
\bibinfo{author}{\bibfnamefont{A.}~\bibnamefont{Bock}} \bibnamefont{et~al.},
  \bibinfo{journal}{Phys. Rev. Lett.} \textbf{\bibinfo{volume}{81}},
  \bibinfo{pages}{534} (\bibinfo{year}{1998}).

\bibitem[{\citenamefont{Isgur and Karl}(1979)}]{Isgur1979}
\bibinfo{author}{\bibfnamefont{N.}~\bibnamefont{Isgur}} \bibnamefont{and}
  \bibinfo{author}{\bibfnamefont{G.}~\bibnamefont{Karl}},
  \bibinfo{journal}{Phys. Rev.} \textbf{\bibinfo{volume}{D19}},
  \bibinfo{pages}{2653} (\bibinfo{year}{1979}).

\bibitem[{\citenamefont{Capstick and Roberts}(2000)}]{Capstick2000}
\bibinfo{author}{\bibfnamefont{S.}~\bibnamefont{Capstick}} \bibnamefont{and}
  \bibinfo{author}{\bibfnamefont{W.}~\bibnamefont{Roberts}},
  \bibinfo{journal}{Prog. Part. Nucl. Phys.} \textbf{\bibinfo{volume}{45}},
  \bibinfo{pages}{S241} (\bibinfo{year}{2000}).

\bibitem[{\citenamefont{Li and Saghai}(1998)}]{Li1998}
\bibinfo{author}{\bibfnamefont{Z.-p.} \bibnamefont{Li}} \bibnamefont{and}
  \bibinfo{author}{\bibfnamefont{B.}~\bibnamefont{Saghai}},
  \bibinfo{journal}{Nucl. Phys.} \textbf{\bibinfo{volume}{A644}},
  \bibinfo{pages}{345} (\bibinfo{year}{1998}).

\bibitem[{\citenamefont{Tiator et~al.}(1994)\citenamefont{Tiator, Bennhold, and
  Kamalov}}]{Tiator1994}
\bibinfo{author}{\bibfnamefont{L.}~\bibnamefont{Tiator}},
  \bibinfo{author}{\bibfnamefont{C.}~\bibnamefont{Bennhold}}, \bibnamefont{and}
  \bibinfo{author}{\bibfnamefont{S.~S.} \bibnamefont{Kamalov}},
  \bibinfo{journal}{Nucl. Phys.} \textbf{\bibinfo{volume}{A580}},
  \bibinfo{pages}{455} (\bibinfo{year}{1994}).

\bibitem[{\citenamefont{Kirchbach and Tiator}(1996)}]{Kirchbach1996}
\bibinfo{author}{\bibfnamefont{M.}~\bibnamefont{Kirchbach}} \bibnamefont{and}
  \bibinfo{author}{\bibfnamefont{L.}~\bibnamefont{Tiator}},
  \bibinfo{journal}{Nucl. Phys.} \textbf{\bibinfo{volume}{A604}},
  \bibinfo{pages}{385} (\bibinfo{year}{1996}).

\bibitem[{\citenamefont{Zhu}(2000)}]{Zhu2000}
\bibinfo{author}{\bibfnamefont{S.-L.} \bibnamefont{Zhu}},
  \bibinfo{journal}{Phys. Rev.} \textbf{\bibinfo{volume}{C61}},
  \bibinfo{pages}{065205} (\bibinfo{year}{2000}).

\bibitem[{\citenamefont{Stoks and Rijken}(1999)}]{Stoks1999}
\bibinfo{author}{\bibfnamefont{V.~G.~J.} \bibnamefont{Stoks}} \bibnamefont{and}
  \bibinfo{author}{\bibfnamefont{T.~A.} \bibnamefont{Rijken}},
  \bibinfo{journal}{Phys. Rev.} \textbf{\bibinfo{volume}{C59}},
  \bibinfo{pages}{3009} (\bibinfo{year}{1999}).

\bibitem[{\citenamefont{Li and Workman}(1996)}]{Li1996}
\bibinfo{author}{\bibfnamefont{Z.-p.} \bibnamefont{Li}} \bibnamefont{and}
  \bibinfo{author}{\bibfnamefont{R.}~\bibnamefont{Workman}},
  \bibinfo{journal}{Phys. Rev.} \textbf{\bibinfo{volume}{C53}},
  \bibinfo{pages}{549} (\bibinfo{year}{1996}).

\bibitem[{\citenamefont{Giannini et~al.}(2001)\citenamefont{Giannini,
  Santopinto, and Vassallo}}]{Giannini2001}
\bibinfo{author}{\bibfnamefont{M.~M.} \bibnamefont{Giannini}},
  \bibinfo{author}{\bibfnamefont{E.}~\bibnamefont{Santopinto}},
  \bibnamefont{and} \bibinfo{author}{\bibfnamefont{A.}~\bibnamefont{Vassallo}},
  \bibinfo{journal}{Eur. Phys. J.} \textbf{\bibinfo{volume}{A12}},
  \bibinfo{pages}{447} (\bibinfo{year}{2001}).

\bibitem[{\citenamefont{Ablikim et~al.}(2006)}]{Ablikim2006}
\bibinfo{author}{\bibfnamefont{M.}~\bibnamefont{Ablikim}} \bibnamefont{et~al.}
  (\bibinfo{collaboration}{BES}), \bibinfo{journal}{Phys. Rev. Lett.}
  \textbf{\bibinfo{volume}{97}}, \bibinfo{pages}{062001}
  (\bibinfo{year}{2006}).

\bibitem[{\citenamefont{Fang}(2006)}]{Fang2006}
\bibinfo{author}{\bibfnamefont{S.-s.} \bibnamefont{Fang}},
  \bibinfo{journal}{Int. J. Mod. Phys.} \textbf{\bibinfo{volume}{A21}},
  \bibinfo{pages}{839} (\bibinfo{year}{2006}).

\bibitem[{\citenamefont{Tryasuchev}(2004)}]{Tryasuchev2004}
\bibinfo{author}{\bibfnamefont{V.~A.} \bibnamefont{Tryasuchev}},
  \bibinfo{journal}{Eur. Phys. J.} \textbf{\bibinfo{volume}{A22}},
  \bibinfo{pages}{97} (\bibinfo{year}{2004}).

\bibitem[{\citenamefont{Mart et~al.}(2004)\citenamefont{Mart, Sulaksono, and
  Bennhold}}]{Mart2004}
\bibinfo{author}{\bibfnamefont{T.}~\bibnamefont{Mart}},
  \bibinfo{author}{\bibfnamefont{A.}~\bibnamefont{Sulaksono}},
  \bibnamefont{and} \bibinfo{author}{\bibfnamefont{C.}~\bibnamefont{Bennhold}}
  (\bibinfo{year}{2004}), \eprint{nucl-th/0411035}.

\bibitem[{\citenamefont{Kelkar and Jain}(1997)}]{Kelkar1997}
\bibinfo{author}{\bibfnamefont{N.~G.} \bibnamefont{Kelkar}} \bibnamefont{and}
  \bibinfo{author}{\bibfnamefont{B.~K.} \bibnamefont{Jain}},
  \bibinfo{journal}{Nucl. Phys.} \textbf{\bibinfo{volume}{A612}},
  \bibinfo{pages}{457} (\bibinfo{year}{1997}).

\bibitem[{\citenamefont{Drechsel et~al.}(1999)\citenamefont{Drechsel, Hanstein,
  Kamalov, and Tiator}}]{Drechsel1999}
\bibinfo{author}{\bibfnamefont{D.}~\bibnamefont{Drechsel}},
  \bibinfo{author}{\bibfnamefont{O.}~\bibnamefont{Hanstein}},
  \bibinfo{author}{\bibfnamefont{S.~S.} \bibnamefont{Kamalov}},
  \bibnamefont{and} \bibinfo{author}{\bibfnamefont{L.}~\bibnamefont{Tiator}},
  \bibinfo{journal}{Nucl. Phys.} \textbf{\bibinfo{volume}{A645}},
  \bibinfo{pages}{145} (\bibinfo{year}{1999}).

\bibitem[{\citenamefont{Davidson et~al.}(1991)\citenamefont{Davidson,
  Mukhopadhyay, and Wittman}}]{Davidson1991}
\bibinfo{author}{\bibfnamefont{R.~M.} \bibnamefont{Davidson}},
  \bibinfo{author}{\bibfnamefont{N.~C.} \bibnamefont{Mukhopadhyay}},
  \bibnamefont{and} \bibinfo{author}{\bibfnamefont{R.~S.}
  \bibnamefont{Wittman}}, \bibinfo{journal}{Phys. Rev.}
  \textbf{\bibinfo{volume}{D43}}, \bibinfo{pages}{71} (\bibinfo{year}{1991}).

\bibitem[{\citenamefont{Dumbrajs et~al.}(1983)}]{Dumbrajs1983}
\bibinfo{author}{\bibfnamefont{O.}~\bibnamefont{Dumbrajs}}
  \bibnamefont{et~al.}, \bibinfo{journal}{Nucl. Phys.}
  \textbf{\bibinfo{volume}{B216}}, \bibinfo{pages}{277} (\bibinfo{year}{1983}).

\bibitem[{\citenamefont{Machleidt et~al.}(1987)\citenamefont{Machleidt,
  Holinde, and Elster}}]{Machleidt1987}
\bibinfo{author}{\bibfnamefont{R.}~\bibnamefont{Machleidt}},
  \bibinfo{author}{\bibfnamefont{K.}~\bibnamefont{Holinde}}, \bibnamefont{and}
  \bibinfo{author}{\bibfnamefont{C.}~\bibnamefont{Elster}},
  \bibinfo{journal}{Phys. Rept.} \textbf{\bibinfo{volume}{149}},
  \bibinfo{pages}{1} (\bibinfo{year}{1987}).

\bibitem[{\citenamefont{Rijken}(2006)}]{Rijken2006}
\bibinfo{author}{\bibfnamefont{T.~A.} \bibnamefont{Rijken}},
  \bibinfo{journal}{Phys. Rev.} \textbf{\bibinfo{volume}{C73}},
  \bibinfo{pages}{044007} (\bibinfo{year}{2006}).

\bibitem[{\citenamefont{Isgur and Karl}(1978)}]{Isgur1978a}
\bibinfo{author}{\bibfnamefont{N.}~\bibnamefont{Isgur}} \bibnamefont{and}
  \bibinfo{author}{\bibfnamefont{G.}~\bibnamefont{Karl}},
  \bibinfo{journal}{Phys. Rev.} \textbf{\bibinfo{volume}{D18}},
  \bibinfo{pages}{4187} (\bibinfo{year}{1978}).

\bibitem[{\citenamefont{Braunschweig et~al.}(1970)}]{Braunschweig1970}
\bibinfo{author}{\bibfnamefont{W.}~\bibnamefont{Braunschweig}}
  \bibnamefont{et~al.}, \bibinfo{journal}{Phys. Lett.}
  \textbf{\bibinfo{volume}{B33}}, \bibinfo{pages}{236} (\bibinfo{year}{1970}).

\bibitem[{\citenamefont{Dewire et~al.}(1971)}]{Dewire1971}
\bibinfo{author}{\bibfnamefont{J.}~\bibnamefont{Dewire}} \bibnamefont{et~al.},
  \bibinfo{journal}{Phys. Lett.} \textbf{\bibinfo{volume}{B37}},
  \bibinfo{pages}{326} (\bibinfo{year}{1971}).

\bibitem[{\citenamefont{Laget}(2005)}]{Laget2005}
\bibinfo{author}{\bibfnamefont{J.~M.} \bibnamefont{Laget}},
  \bibinfo{journal}{Phys. Rev.} \textbf{\bibinfo{volume}{C72}},
  \bibinfo{pages}{022202} (\bibinfo{year}{2005}).

\bibitem[{\citenamefont{Rodrigues et~al.}(2008)}]{Rodrigues2008}
\bibinfo{author}{\bibfnamefont{T.~E.} \bibnamefont{Rodrigues}}
  \bibnamefont{et~al.}, \bibinfo{journal}{Phys. Rev. Lett.}
  \textbf{\bibinfo{volume}{101}}, \bibinfo{pages}{012301}
  (\bibinfo{year}{2008}).

\bibitem[{\citenamefont{Sibirtsev et~al.}(2010)\citenamefont{Sibirtsev,
  Haidenbauer, Krewald, and Meissner}}]{Sibirtsev2010}
\bibinfo{author}{\bibfnamefont{A.}~\bibnamefont{Sibirtsev}},
  \bibinfo{author}{\bibfnamefont{J.}~\bibnamefont{Haidenbauer}},
  \bibinfo{author}{\bibfnamefont{S.}~\bibnamefont{Krewald}}, \bibnamefont{and}
  \bibinfo{author}{\bibfnamefont{U.~G.} \bibnamefont{Meissner}},
  \bibinfo{journal}{Eur. Phys. J.} \textbf{\bibinfo{volume}{A44}},
  \bibinfo{pages}{169} (\bibinfo{year}{2010}).

\bibitem[{\citenamefont{Koniuk and Isgur}(1980)}]{Koniuk1980}
\bibinfo{author}{\bibfnamefont{R.}~\bibnamefont{Koniuk}} \bibnamefont{and}
  \bibinfo{author}{\bibfnamefont{N.}~\bibnamefont{Isgur}},
  \bibinfo{journal}{Phys. Rev.} \textbf{\bibinfo{volume}{D21}},
  \bibinfo{pages}{1868} (\bibinfo{year}{1980}).

\end{thebibliography}
\end{document}